\documentclass{JHEP3}

\usepackage{amsmath}
\usepackage{amssymb}
\usepackage{graphicx}
\usepackage{slashed}

\newcommand{\bZ}{\mathbb{Z}}
\newcommand{\bC}{\mathbb{C}}

\newcommand{\cN}{\mathcal{N}}

\newcommand{\cD}{\mathcal{D}}

\newcommand{\ov}{\overline}
\newcommand{\ket}[1]{\left|#1\right\rangle}
\newcommand{\vev}[1]{\left\langle#1\right\rangle}

\title{Branes and instantons intersecting at angles}

\author{Mirjam Cveti\v{c}, I\~naki Garc\'ia-Etxebarria and Robert Richter\\
  Department of Physics and Astronomy, University of Pennsylvania,\\
  Philadelphia, PA 19104-6396, USA\\
  E-mail: \email{cvetic@cvetic.hep.upenn.edu},
  \email{inaki@sas.upenn.edu}, \email{rrichter@sas.upenn.edu}}

\abstract{We study in detail the system of D6 branes and euclidean
  D2-brane instantons intersecting at angles in type IIA string
  theory. We find that in the absence of orientifolds the system does
  not contribute to the low energy superpotential, in agreement with
  expectations based on effective field theory arguments. We also
  comment on the implications of our results for dual string theory
  pictures.}

\preprint{UPR-1207-T}

\begin{document}

\section{Introduction}

In the last few years there has been growing interest in
non-perturbative effects coming from euclidean D-brane instantons
\cite{Becker:1995kb}, mainly due to the realization
\cite{Blumenhagen:2006xt,Ibanez:2006da,Florea:2006si} that they could
be used for solving some longstanding difficulties in string model
building coming from perturbative symmetries. We refer the reader to
\cite{Blumenhagen:2009qh} for a survey of the recent developments.

\medskip

Of central importance in the study of instanton effects is the issue
of fermionic zero modes. They are Grassmann variables, and integration
over their measure makes the contribution of the instanton to a given
observable vanish unless they are properly saturated by explicit
insertions. This severely constrains the quantities a given instanton
will contribute to.

\medskip

In this paper we will be mainly interested in contributions to the low
energy effective superpotential $W$ of a $\cN=1$ theory. Our focus
will be on four dimensional $\cN=1$ theories realized as Calabi-Yau
compactifications of type II string theory in the presence of D$(3+p)$
branes, and for the most part we will be dealing with type IIA in the
presence of D6 branes. We want to understand instanton effects coming
from a single euclidean D2 brane wrapping some special Lagrangian
cycle in the internal manifold, and localized in the four Minkowski
directions.

\medskip

Fermionic zero modes of the instanton come from open strings going
from the instanton to itself, or to background D6 branes. Because of
their charge under the gauge $U(1)$ of the D6 branes, zero modes of
the second kind are commonly called {\em charged}, while those of the
first kind are called {\em neutral}. The biggest constraint in model
building with instantons comes from the neutral zero modes, they are
the ones we will be trying to understand in this paper.

\medskip

The issue with neutral zero modes in this class of compactifications
is the following: despite the fact that the background preserves four
supercharges, away from the background branes and orientifolds (and in
the absence of flux, see below) there are eight conserved
supercharges, since locally we are dealing with a Calabi-Yau
compactification of type II. Our D-brane instantons are 1/2 BPS
objects, so they break four of the eight supersymmetries. The broken
supersymmetries will be nonlinearly realized as goldstinos in the
worldvolume of the instanton. We will denote these zero modes
$\theta_\alpha$ and $\ov\tau_{\dot\alpha}$. The two $\theta_{\alpha}$
goldstinos are expected for a superpotential: one gets the coupling
in the Lagrangian from the term going with $\theta^2$ in the expansion
of the superpotential superfield. In string theory this is realized as
a disk diagram with explicit insertions of two $\theta_{\alpha}$ modes
(see \cite{Blumenhagen:2006xt}, for example). On the other hand, the
$\ov\tau_{\dot\alpha}$ should be removed from the spectrum if we want
to generate a superpotential.

\medskip

There are various known mechanisms for saturating or removing the
$\ov\tau$ modes. The simplest one is to put one or more space-filling
branes on top of the instanton. In this case, reviewed in
section~\ref{sec:gauge-review}, the instanton is interpreted as a
gauge instanton in the theory of the space-filling brane.\footnote{The
usual gauge effects coming from instantons are reproduced in this
way. As an example, \cite{Akerblom:2006hx} gives a concrete
realization of the ADS superpotential using D-brane instantons.} The
case with a single space-filling brane is slightly special, since it
does not admit a gauge theory interpretation, but the lifting of
$\ov\tau$ modes still takes place
\cite{Aganagic:2007py,GarciaEtxebarria:2007zv,Petersson:2007sc,GarciaEtxebarria:2008iw,Ferretti:2009tz}. When
there are orientifolds in the system, instantons wrapping cycles
invariant under the orientifold action and such that the gauge
symmetry on their worldvolume is $O(1)$ (the so-called \emph{$O(1)$
  instantons}) are also interesting. In this case the orientifold
projects out the undesired $\ov \tau$ modes
\cite{Argurio:2007qk,Ibanez:2007rs,Argurio:2007vqa,Bianchi:2007wy} and
superpotential contributions can be generated. Along similar lines,
multi-instanton configurations which can recombine into a $O(1)$
instanton can saturate all zero modes and thus contribute to the
superpotential \cite{Blumenhagen:2007bn,
  GarciaEtxebarria:2007zv,Cvetic:2008ws,GarciaEtxebarria:2008pi}.  The
last possibility is to consider the instanton in the presence of
background flux
\cite{Marolf:2003ye,Marolf:2003vf,Kallosh:2005gs,Martucci:2005rb,Bergshoeff:2005yp,Park:2005hj,Bandos:2005ww,Lust:2005cu,Tsimpis:2007sx,Schulgin:2007zz,Blumenhagen:2007bn,GarciaEtxebarria:2008pi,Billo':2008pg,Billo':2008sp,Uranga:2008nh}. In
most known scenarios the fluxes one has to introduce in order to lift
$\ov\tau$ modes necessarily break supersymmetry
\cite{Blumenhagen:2007bn,GarciaEtxebarria:2008pi,Billo':2008pg,Billo':2008sp,Uranga:2008nh},
a fact that can be understood most easily from the 4d effective field
theory point of view \cite{GarciaEtxebarria:2008pi,Uranga:2008nh}.

\medskip

The focus of this paper is a careful analysis of a system composed of
a D6 brane and a rigid D-brane instanton intersecting at angles in the
internal space, in the absence of orientifolds or fluxes.\footnote{A
  related system has been recently studied in
  \cite{Heckman:2008es,Marsano:2008py}.} This system is of particular
interest in the study of D-brane instanton effects in particle physics
models. Our key result is that in this case the $\ov\tau_{\dot\alpha}$
modes are not lifted, and thus no superpotential is generated by the
D-brane instantons. This negative result has important implications
for the particle physics model building, since it excludes such
ubiquitous configurations from generating superpotential terms. In the
following we will elaborate in detail on the origin of this rather
strong result.

\medskip

The main tool we will use in our analysis is a microscopic CFT
calculation, although we will also discuss some general effective
field theory arguments forbidding the generation of the
superpotential. This is the main reason why we have chosen to work in
type IIA: close to the point(s) where the instanton and the brane
intersect one has a good CFT description of the system, and the fact
that the $\ov\tau$ modes  are not lifted can be shown
convincingly. The type IIA picture is also related by duality to
backgrounds in type IIB, and we comment on implications of our result
for such systems.

\medskip

In our CFT analysis we will restrict to the oscillator modes that
become massless when the D6 gets aligned with the instanton. More
precisely, in terms of its dependence on the angles $\theta_i$
parameterizing the local rotation between the instanton and the brane,
the mass of any oscillator mode is of the form (in string units):
\begin{equation}
  m = c + \sum_{i=1}^3 a_i \theta_i
\end{equation}
with $c$ and $a_i$ being some constants depending on the state. We will
restrict to modes with $c=0$. Modes with $c\neq 0$ are massive and
genuinely stringy, and we believe that they will not affect the
results of our analysis. The main reason for believing this is that
similar modes are present in the gauge instanton case, where they do
not modify the analysis in any substantial way.

\medskip

We will also restrict to the vicinity of a single intersection, which
can be modelled by a system of branes at angles in flat space. The
general configuration of interest in model building will have D6
branes wrapping special Lagrangian manifolds, possibly intersecting at
more than one point. Nevertheless, the result of the analysis still
applies simply because $\ov\tau$ is not saturated in any intersection,
and thus cannot be saturated globally.

\subsubsection*{A holomorphy argument, and the importance of tachyons}
We will be mostly interested in the case where the brane and the
instanton share some supersymmetries. Nevertheless, the fact that
there are some common supersymmetries is only true at very special
points in complex structure moduli space, and in a generic point in
complex structure moduli space the brane and the instanton will be
misaligned, preserving no common supersymmetries. By holomorphy of the
generated low energy superpotential, this implies
\cite{GarciaEtxebarria:2007zv,GarciaEtxebarria:2008pi} that no
superpotential can be generated anywhere in complex structure moduli
space, including the points where the brane and the instanton share
some supersymmetries.\footnote{For related considerations, we also
  refer the reader to the recent work \cite{Collinucci:2009nv}, where
  the global behavior of instanton contributions to the superpotential
  has been linked to the topological string.}

One worry with this argument is that away from the supersymmetric
point in moduli space, the instanton might dissolve into the brane,
restoring supersymmetry. As discussed microscopically in
\cite{Blumenhagen:2007bn} and applied for global considerations in
\cite{GarciaEtxebarria:2007zv,GarciaEtxebarria:2008pi} something
similar happens in the case where we have a two-instanton process
where the two instantons misalign: the system will still give us a
superpotential contribution if the two misaligned instantons can
recombine into a single BPS instanton with the proper structure of
zero modes. In the context of the system discussed in this paper, such
a recombination process would be signalled by the appearance of a
tachyonic open string between the instanton and the brane, coming from
the bosonic modes we will denote $\omega_{\dot\alpha}$ in the
following.

It is easy to argue that there is no such tachyon; the argument goes
as follows. We will see in section~\ref{sec:cft} that the bosonic
modes have a positive mass in the supersymmetric case (see also
\cite{Blumenhagen:2006xt}). Generic infinitesimal deformations of the
angles make the system non-supersymmetric, while keeping the mass of
$\omega_{\dot\alpha}$ bigger than zero. In order for the misaligned
instanton to contribute, it must have a tachyon at each
non-supersymmetric point, so we conclude that recombination is not
possible, and the instanton does not contribute.

Remarkably, there are systems where there {\em is} a tachyon between
the brane and the instanton, and the misaligned instanton just
dissolves into the brane, restoring supersymmetry. As an illustration
of this effect, in section~\ref{sec:noncommutative} we discuss the
non-commutative gauge instanton.

\subsubsection*{The microscopic analysis}

The previous macroscopic argument is powerful and general, but it does
not show what goes wrong microscopically. Let us discuss here one such
puzzle arising from the microscopic point of view.

Misaligned with respect to the brane or not, there are always 4
neutral fermionic zero modes in the instanton worldvolume due to the
fact that it is a 1/2 BPS object of the Calabi-Yau background (we are
assuming that there is no orientifold around). In the case of the
gauge instanton, two of the four zero modes correspond to
supersymmetries broken by both the instanton and the brane. This
partial supersymmetry breaking by the background brane appears as an
effective coupling in the instanton worldvolume lifting these extra
``pseudo-goldstinos'' (we review all this more carefully in
section~\ref{sec:gauge-review}).

In the case where the instanton and the brane are intersecting and
misaligned there are no shared supersymmetries, so one might expect
four genuine goldstinos not lifted by any interactions. Thus, as
mentioned above, the results of \cite{GarciaEtxebarria:2008pi} are
expected to apply. Nevertheless, it is not hard to see that there are
couplings in the instanton action similar to those that lift the
``pseudo-goldstinos'' in the gauge instanton case (we will show this
using CFT techniques in section~\ref{sec:cft}), only this time
 $\ov\tau_{\dot\alpha}$ couples to massive modes. One might still
wonder what is the effect of these couplings.

More precisely, in \eqref{eq:instD-brane-action} we will find the
following action for the modes going from the instanton to the brane:
\begin{equation}
  S^{E2-D6}=   m_{\ov \mu^A}\, \ov \mu^{A} \, \mu'_{A} + m_{\mu^B}
  \,\mu^{B} \, \ov \mu'_{B} + m_{\omega} \,  \omega \, \ov \omega +
  \ov \tau\, (\ov \omega\, \mu^0 + \ov \mu^0 \, \omega) + i
  \vec{D}\cdot
  \omega \,\vec{\sigma}\,\bar\omega
\end{equation}
where $\ov\tau_{\dot\alpha}$, $\mu^A$, $\mu'_A$, $\ov\mu^A$ and
$\ov\mu'_A$ are the Grassmann variables we have to saturate in the
measure, $\ov\omega^{\dot\alpha}$ is the lightest bosonic modes from
the instanton to the brane, and $D$ is the D-term auxiliary
field. Note the presence of the primed modes $\mu'$, $\ov\mu'$, which
are not present in the gauge instanton case. They are required in
order to give masses to $\mu$ and $\bar\mu$. As we will describe in
section~\ref{sec:effective-action}, this action is compatible with the
supersymmetries preserved by the system, and in fact its form is
highly constrained by these same supersymmetries. It is also easy to
see (we show this in detail in section~\ref{sec:saturation}) that with
such interactions we cannot saturate all Grassmann integrations: if we
bring down the second coupling in order to saturate $\ov\mu'$, then we
also saturate $\mu$, and we cannot use the first term to saturate
$\ov\tau$. We believe that our CFT analysis in section~\ref{sec:cft}
settles conclusively this and related issues arising from the
microscopic point of view.

\medskip

This paper is organized as follows. In section~\ref{sec:gauge-review}
we review some background material on the embedding of gauge
instantons in string theory that might be useful for comparison with
the intersecting brane case. Section~\ref{sec:cft} contains the core
of this work, a CFT analysis of the brane-instanton
system. Section~\ref{sec:dual-pictures} contains some comments on
systems related by duality. Finally, section~\ref{sec:noncommutative}
contains a short discussion of the non-commutative instanton as an
illustration of a system where recombination is allowed.

\section{A short review of gauge instantons in string theory}
\label{sec:gauge-review}

Before starting the analysis of the somewhat exotic intersecting brane
and instanton system in section~\ref{sec:cft}, let us review here some
important features of the gauge instanton case. As we will see, there
are some crucial differences between both cases, but also some
important parallels that make it worth to keep the gauge case in
mind. Here we will just discuss the main points of interest to us, we
refer the reader to \cite{Green:2000ke,Dorey:2002ik} for much more
detailed discussions.

\medskip

In many ways the nicest system where we can discuss gauge instantons
in string theory is $\cN=4$ $SU(N)$ SYM, which can be engineered
simply by putting $N$ coincident D3 branes in flat space. The gauge
instantons of this theory appear as D$(-1)$ branes dissolved on the
worldvolume of the D3. The moduli space of instantons in field theory
appears as the Higgs branch of the moduli space of vacua of the theory
living on the worldvolume of the D$(-1)$ brane
\cite{Witten:1995gx}. Although it is possible to work directly in the
$0+0$ dimensional worldvolume theory of the D$(-1)$ instanton, and in
fact for the intersecting case we will be forced to work in this way,
for the $\cN=4$ theory it is simpler to work in the D5-D9 system
instead, and at the end dimensionally reduce the results along the 6
common directions. This makes the effect of the supersymmetry clearer,
and the interpretation of the fields in the D$(-1)$ worldvolume more
transparent. Since they do not play a role in the intersecting case,
we will set the fields living on the D9 to zero.

\subsubsection*{The instanton worldvolume theory}

Let us describe the worldvolume theory on the D5. As a matter of
convention, we will denote the spinorial indices in the 6d theory by
$A,B,\ldots=1\ldots 4$, and the vector indices by $m,n,\ldots=1\ldots
6$. The D5 preserves 16 supersymmetries of the background, in the
usual 6d notation we have $\cN=(1,1)$. The background D9 will break
down this further to $\cN=(0,1)$, so we will express the multiplets in
this language. On the D5 there is a vector multiplet $(\chi_m,
\ov\tau^A_{\dot\alpha},D_i)$. Here $\chi_m$ is the gauge connection
on the D5, $D_i$ is a triplet of auxiliary D-terms, and
$\ov\tau^A_{\dot\alpha}$ is the left half of the reduction of the 10d
Majorana-Weyl gaugino, which decomposes under $SO(6)\times SO(4) \sim
SU(4)\times SU(2)_R\times SU(2)_L$ as:
\begin{equation}
  \mathbf{16} \longrightarrow (4,2,1) + (\bar 4, 1, 2)\,\,.
\end{equation}
The right half $\theta^{\alpha}_A$ of the decomposition appears in an
adjoint hypermultiplet with four real bosonic components $x_\mu\sim
x_{\alpha\dot\alpha}$. These two multiplets of $\cN=(0,1)$ join into
the vector multiplet of the $\cN=(1,1)$ symmetry on the D5.

The strings going from the D5 to the D9 break down $\cN=(1,1)$ to
$\cN=(0,1)$. They give a single hypermultiplet in the fundamental,
with two complex bosonic components $\omega_{\dot\alpha}$ and one Weyl
fermion $\mu^A$.

There is a Lagrangian for the theory living on the D5, it can be
written as the sum of three terms:
\begin{equation}
  \label{eq:N4-inst}
  S_{inst} = \int d^6\xi\,\mathrm{Tr}\left( S_{gauge} + S^{(1,1)}_{matter} +
    S^{(0,1)}_{matter} \right)
\end{equation}
with $\xi$ the coordinates in the D5. The first term encodes the gauge
dynamics:
\begin{equation}
  S_{gauge} = \frac{1}{2} F^2 - i \ov\tau
  \slashed{\cD}\ov\tau + \frac{1}{2}\vec{D}^2
\end{equation}
where we are using the reality condition on $\ov\tau$,\footnote{This
  condition is inherited from the Majorana-Weyl condition on the
  original 10d spinor. It reads $(\ov\tau)^{\dagger} = \bar\Sigma_0
  \ov\tau$, with $\bar\Sigma$ the gamma matrix for anti-chiral 6d Weyl
  spinors.} and $\slashed{\mathcal{D}}$ is the covariant derivative
contracted with the 6d chiral gamma matrices $\bar\Sigma$:
$\slashed{\cD}=\bar\Sigma_m\cD_m$. (In these expressions we will
sometimes omit some obvious index contractions for clarity, so
$\ov\tau \slashed{\cD}\ov\tau$ stands for
$\ov\tau^A_{\dot\alpha}\slashed{\cD}_{AB}\ov\tau^{B\dot\alpha}$, etc.)

The other two terms encode the dynamics of the hypermultiplets. For
the hypermultiplet $(x,\theta)$ in the adjoint of the D5 gauge group
we have:
\begin{equation}
  S^{(1,1)}_{matter} = \cD_m x_{\alpha\dot\alpha} \cD^m
  x^{\alpha\dot\alpha} -
  i\theta
  \slashed{\cD} \theta -i [\theta^\alpha, x_{\alpha\dot\alpha}] \ov\tau^{\dot\alpha} + i \vec{D}\cdot
  \mathrm{tr}(x\vec{\sigma}x)
\end{equation}
with $\vec{\sigma}$ the Pauli sigma matrices. Similarly for the
hypermultiplet in the fundamental:
\begin{equation}
  S^{(0,1)}_{matter} = \cD_m \bar\omega^{\dot\alpha} \cD^m
  \omega_{\dot\alpha} - i \bar\mu \slashed{\cD} \mu -
  i(\bar\mu^A\omega_{\dot\alpha} +
  \bar\omega_{\dot\alpha}\mu^A)\ov\tau^{\dot\alpha}_A + i \vec{D}\cdot
  \omega_{\dot\alpha}\vec{\sigma}^{\dot\alpha\dot\beta}\bar\omega_{\dot\beta}
\end{equation}

The theory on the D$(-1)$ can be easily obtained by dimensionally
reducing (\ref{eq:N4-inst}), simply by discarding the derivatives. We
will also reduce to the case of a single D3 and a single D$(-1)$, so
all the commutators vanish. The fields now become zero modes of the
instanton: $\chi_m$ parameterizes the position of the D(-1) on the
internal ${\bf R}^6$ (i.e., it parameterizes motion away from the D3),
$x_\mu$ is the position of the instanton in the ${\bf R}^4$ parallel
to the D3, $\omega$ parameterizes the Higgs branch of the instanton,
where it gets dissolved into the D3 and admits a classical
interpretation as a gauge instanton, and $\ov\tau$ and $\theta$ are
the Goldstinos of the supersymmetries broken by the instanton. The
resulting action (restoring some constant factors we have disregarded
in the analysis above) is given by \cite{Green:2000ke}:
\begin{equation}
  \label{eq:D(-1)-action}
  S_{inst} = -2\pi i\tau + \frac{1}{g_0^2}\vec{D}^2 + \chi^2W_0 -
  \frac{2i}{\sqrt{8}}\bar\mu^A\mu^B\Sigma^m_{AB} \chi_m + i
  \vec{D}\cdot \vec{W} +
  i\ov\tau_A^{\dot\alpha}(\bar\mu^A \omega_{\dot\alpha} +
  \mu^A\bar\omega_{\dot\alpha})
\end{equation}
where we have introduced
\begin{equation}
  g_0^2 = 4\pi(4\pi^2\alpha')^{-2} g_s \qquad ; \qquad \tau = C_0 - \frac{i}{g_s}
\end{equation}
with $g_s$ the string coupling, $C_0$ the RR 0-form of type IIB, and
\begin{equation}
  \label{eq:Wc}
  W_0 = \bar\omega \omega \qquad;\qquad W^c = \bar\omega
  \bar\sigma^{\mu\nu}\omega \bar\eta^c_{\mu\nu}
\end{equation}
with $\bar\eta$ the anti-self-dual 't Hooft symbol mapping
anti-self-dual tensors to the adjoint of $SU(2)$ (see Appendix~A of
\cite{Dorey:2002ik} for explicit expressions, which we will not need
here). We also have the relation:
\begin{equation}
  W_0^2 = \sum_c (W^c)^2
\end{equation}

Notice that in the field theory limit of an instanton on its Higgs
branch\footnote{This limit is somewhat subtle. We need to take
  $\alpha'\rightarrow 0$, keeping the string coupling small so that
  the D3 theory becomes weakly coupled SYM. We also need to restrict
  to vevs for $\omega$ such that we stay away from the small instanton
  singularity.} the second term in (\ref{eq:D(-1)-action}) disappears
(since $g_0\rightarrow \infty$) and $\vec{D}$ becomes a Lagrange
multiplier, which implements the bosonic ADHM constraints
\cite{Atiyah:1978ri}. We can use the last term in
(\ref{eq:D(-1)-action}) in order to saturate $\ov\tau$, the resulting
insertion in the path integral implements the so-called fermionic ADHM
constraints $(\bar\mu^A \omega_{\dot\alpha} +
\mu^A\bar\omega_{\dot\alpha})=0$. In section~\ref{sec:backreaction} we
will be interested in the behavior of the small instanton limit, so we
will keep the second term in \eqref{eq:D(-1)-action} around.

\subsubsection*{Supersymmetry transformations}

The supersymmetry transformations of the zero modes under the unbroken
supersymmetries $\bar\xi_{\dot\alpha A}$ can be found either by a CFT
computation \cite{Billo:2002hm,Green:2000ke}, similar to the one we
will do in section~\ref{sec:cft}, or by field theory considerations
\cite{Dorey:2002ik}. The result is:
\begin{align}
  \delta x_{\alpha\dot\alpha} & = i \bar \xi_{\dot\alpha
    A}\theta^{A}_{\alpha} & \delta \theta^A_{\alpha} & = 0\\
  \delta \chi_m & = i \Sigma_m^{AB} \bar\xi_A \ov\tau_B
  & \delta\ov\tau_A & = \vec{D}\cdot \vec{\sigma} \bar\xi_{A} \\
  \delta \omega_{\dot\alpha} & = i \bar\xi_{\dot\alpha A}\mu^A &
  \delta\mu^A & = 0
\end{align}
where we have left out some terms that will not be relevant for our
discussion (the full expressions can be found in section~X.3.1 of
\cite{Dorey:2002ik}).

% We do not want the footnote to appear in the table of contents.
\newcommand{\zzsectitle}{Backreaction and the Coulomb branch of $\cN=4$
  instantons}
\subsection[\zzsectitle]{\zzsectitle\footnote{We will not use the
    results of this section in the rest of the paper, so it can be
    skipped in a first reading. We include it here as an interesting
    side remark.}}
\label{sec:backreaction}

As we discussed in the introduction, lifting the neutral zero modes of
stringy instantons is somewhat involved in general type II
compactifications. The main reason is that the compactification
locally preserves 8 supersymmetries, and the instanton breaks 4 of
these, so it necessarily has 4 goldstinos.

\medskip

It is interesting to consider what happens when the backreaction of
the background branes is taken into account. Typically, this analysis
is complicated by the absence of explicit expressions for the
backreacted backgrounds, and the difficulty of applying CFT techniques
to general curved backgrounds.\footnote{Nevertheless, there has been
  some recent progress in the analysis of instantons in the presence
  of fluxes from the point of CFT
  \cite{Billo:2006jm,Billo':2008sp,Billo':2008pg}.} In this section we
would like to argue that the D3-D$(-1)$ system in its Coulomb branch
(defined by $\chi_m\neq 0$) provides a particularly simple toy example
in which the backreaction of the D3 lifts all the
$\ov\tau^A_{\dot\alpha}$ modes.

\medskip

In the Coulomb branch of the instanton the modes going from the D3 to
the D$(-1)$ (the hypermultiplet in the fundamental, and its stringy
excitations) acquire a mass growing with $\chi_m$. Our strategy will
be to integrate these modes out explicitly, finding an effective
instanton action on the Coulomb branch depending just on
$\chi_m$. Schematically (we make this more precise below):
\begin{equation}
  \label{eq:eff-S-coulomb}
  e^{-S_{inst}(\chi_m)} \sim \int d[\omega,\mu,\ldots]\, e^{-S_{inst}(\chi_m,\,\omega,\,\mu,\,\ldots)}
\end{equation}
where the measure includes all the massive modes. We will find out
that the integral on the right side saturates the integral over
$\ov\tau$ for any $\chi_m$. In terms of string diagrams, we want to
compute the cylinder diagram with one boundary on the D3 and the other
on the D$(-1)$, with various insertions of $\ov\tau$ on the D$(-1)$
boundary. This is a complicated problem, and we are just interested
here in the qualitative behavior of zero mode lifting, so let us
restrict ourselves to two particularly simple limits in which the
cylinder contribution can be easily obtained.

\medskip

The first limit we want to consider is $\chi_m$ much smaller than the
string scale. In this limit we can restrict to the open string modes
of lowest mass $\omega$ and $\mu$, and just plug
(\ref{eq:D(-1)-action}) into (\ref{eq:eff-S-coulomb}).\footnote{A
  similar computation, with a different motivation, was performed in
  \cite{Green:2000ke}. We will follow this reference in doing our
  computation.} It will be convenient to integrate out explicitly the
D-terms first. If we do this we get:
\begin{equation}
  S_{inst} = -2\pi i\tau + \frac{g_0^2}{4}(W^c)^2 + \chi^2W_0 -
  \frac{2i}{\sqrt{8}}\bar\mu^A\mu^B\Sigma^m_{AB} \chi_m +
  i\ov\tau_A^{\dot\alpha}(\bar\mu^A \omega_{\dot\alpha} + \mu^A\bar\omega_{\dot\alpha})\,\,.
\end{equation}
Furthermore, in doing this integration we pick up a prefactor of
$g_0^3$ in the measure of integration of zero modes coming from the
$1/g_0^2$ coefficient of $\vec{D}^2$ in (\ref{eq:D(-1)-action}). The
integral to perform is now:
\begin{equation}
  g_0^3\int d[\omega,\bar\omega,\mu,\bar\mu]\,\exp\left[2\pi i\tau - \frac{g_0^2}{4}(W^c)^2 - \chi^2W_0 +
    \frac{2i}{\sqrt{8}}\bar\mu^A\mu^B\Sigma^m_{AB} \chi_m -
    i\ov\tau_A^{\dot\alpha}(\bar\mu^A \omega_{\dot\alpha} + \mu^A\bar\omega_{\dot\alpha})\right]\,\,.
\end{equation}
The first term in the exponential does not depend on the massive
modes, so we will ignore it in the following. We want to focus on the
terms that saturate as many $\ov\tau$ modes as possible, so let us
saturate all $\mu,\bar\mu$ integrations by bringing down the last term
eight times. We are left with:
\begin{equation}
  \label{eq:omega-integral}
  (\ov\tau)^8 \int d[\omega,\bar\omega]\, W_0^4\exp\left[-\frac{g_0^2}{4}(W^c)^2 - \chi^2W_0\right]\,\,.
\end{equation}
Here $(\ov\tau)^8$ is shorthand for a coupling that saturates all
$\ov\tau$ modes, with the normalization:
\begin{equation}
  \int d^8\ov\tau\,(\ov\tau)^8 = 1\,\,.
\end{equation}
We are left with the integral over $\omega,\bar\omega$. At this point
it is very convenient to change variables from $\omega$ to $W^c$:
\begin{equation}
  \int d[\omega,\bar\omega] = 2\pi \int \frac{dW^1dW^2dW^3}{W_0}\,\,.
\end{equation}
Since our integral (\ref{eq:omega-integral}) depends just on the
radial coordinate $W_0$, we can perform the angular part of the
integration, and we end up with:
\begin{equation}
  \begin{split}
  e^{-S_{inst}(\chi_m)} &= 8\pi^2g_0^3 (\ov\tau)^8
  \int_0^{\infty}dW_0\,W_0^5 \exp\left[-\frac{g_0^2}{4}(W_0)^2 -
    \chi^2W_0\right]\\
  & = 8\pi^2g_0^{-3} (\ov\tau)^8 \int_0^{\infty}dw\,w^5
  \exp\left[-\frac{1}{4}(w)^2 -
    w\frac{\chi^2}{\sqrt{g_0}}\right]\\
  & = 8\pi^2g_0^{-3} (\ov\tau)^8\left(32 t^4 + 144 t^2 + 64 - (32t^5
    + 160 t^3 + 120t)\sqrt{\pi}e^{t^2}\mathrm{erfc}(t)\right)
  \end{split}
\end{equation}
where we have introduced the adimensional variables $w\equiv g_0 W_0$
and $t\equiv\chi^2/\sqrt{g_0}$, and $\mathrm{erfc}(t)$ denotes the
complementary error function \cite{abramowitz+stegun}. While this is
perhaps not a particularly transparent expression, it does show what
we want, namely that there is a lifting of the $\ov\tau$ modes in the
Coulomb branch of the instanton. This lifting is maximized at the
origin of the Coulomb branch, and decreases fast as $\chi^2$
increases.

\medskip

Let us now consider the opposite limit, where $\chi_m$ is much larger
than the string scale. Here the cylinder can be computed in the closed
string channel, with a closed string mode propagator connecting a disk
with the boundary on the D$(-1)$ and a disk with the boundary on the
D3. In this regime we can replace the D3 by its backreaction, and just
compute the disk tadpole for the D$(-1)$ with a closed string mode
inserted on it. The D3 sources the RR 4-form $C_4$ and the metric
$G_{mn}$, so we need the couplings of a disk with boundary on the
D$(-1)$, an insertion of $C_4$ or $G_{mn}$ on the bulk of the disk,
and various insertions of $\ov\tau$ on the boundary. This
computation, while easier than the whole cylinder diagram, is still
quite challenging. We will not present the final result, but let us
point out that the requisite couplings are present. It was argued in
\cite{Green:1997tv} that there is a coupling between $G_{mn}$ and four
goldstinos of the instanton:
\begin{equation}
  \label{eq:D(-1)-G}
  S_{inst} \sim \ldots + \partial_m \partial_n G_{\rho\sigma}
  (\bar\Theta\Gamma^{m\rho\nu}\Theta)(\bar\Theta\Gamma^{n\sigma\nu}\Theta)
\end{equation}
where $\Theta$ is the Majorana-Weyl spinor encoding the neutral
fermionic zero modes of the D$(-1)$, and $\Gamma^{m\mu\nu}$ is the
antisymmetrized product of three $\Gamma$ matrices. The greek indices
take values in the 10 euclidean dimensions. The coupling to $C_4$ can
be obtained T-dualizing the coupling of a D$0$ brane to $C_5$ obtained
in \cite{Millar:2000ib}. This gives the following coupling:
\begin{equation}
  S_{inst} \sim \ldots + \partial_m \partial_n C^{(4)}_{\mu\nu\rho\sigma}
  (\bar\Theta\Gamma^{m\mu\nu}\Theta)(\bar\Theta\Gamma^{n\rho\sigma}\Theta)
\end{equation}

\subsubsection*{Extension to theories with less supersymmetry}

One very interesting aspect of the previous $\cN=4$ analysis is that
it points towards a hitherto unnoticed mechanism for lifting neutral
zero modes in $\cN=1$ cases. As mentioned in the beginning of this
section, doing the analysis explicitly is difficult, but we can easily
give a plausibility argument based on the $\cN=4$ case just discussed.

The previous discussion holds without big modifications in any theory
where stringy instantons have both a Coulomb and Higgs branch,
connected through the small instanton limit. One such theory is
$\cN=2$ $SU(N)$ SYM, which we can engineer in string theory in the
following way. Take a local geometry given by ${\bf R}^6$ times a
(resolved) $A_1$ degeneration of K3:
\begin{equation}
  x^2 + y^2 + w^2 = 0,
\end{equation}
and wrap $N$ D5-branes on the resolved ${\bf S}^2$ times some ${\bf
  R}^4\subset {\bf R}^6$. Similarly to the $\cN=4$ case, gauge
instantons in this theory have a Coulomb branch, parameterizing the
${\bf R}^2$ where the D5 is pointlike, and represented on the field
theory of the instanton by an adjoint hypermultiplet. As in the
$\cN=4$ case, the whole Coulomb branch of the instanton will
contribute to the theory on the D5.

\medskip

Let us now deform the previous space by fibering the $A_1$ singularity
over the ${\bf R}^2$ plane, which we parameterize by $z$
\cite{Cachazo:2001gh}:
\begin{equation}
  x^2 + y^2 + w^2 = f(z)^2
\end{equation}
This geometry breaks the supersymmetry down to $\cN=1$, and in
particular lifts the Coulomb branch to a set of isolated points, the
set of solutions to $f(z)=0$. Close to each of these points, for
sufficiently generic $f(z)$, we have a resolved conifold
singularity. Branes wrapped on the ${\bf S}^2$ of these resolved
conifolds are BPS, and in particular we will generically have isolated
instantons not on top of any brane or orientifold. We can make $f(z)$
as small as we want, so we expect that the basic mechanism that lifted
zero modes in the $\cN=4$ case still applies, and a superpotential is
generated.

\medskip

The utility of this mechanism for stabilizing K\"ahler moduli is
limited by the fact that the instanton is homologous to the gauge
brane, so there will be a contribution to the potential only if there
is gaugino condensation already (and our mechanism is rather
subleading in this case). Nevertheless we find it of some theoretical
interest, and worthy of further study.

\section{Conformal field theory analysis}
\label{sec:cft}

In this chapter we perform a detailed analysis of the instanton modes
and their interactions using conformal field theory techniques. For
concreteness we perform this analysis in type IIA, where the relevant
instantons are euclidean D2 branes wrapping a three-cycle $\Sigma$
away from the orientifold plane in the internal manifold (and thus
point-like in four-dimensional space-time). Generically, such an
isolated $U(1)$ instanton exhibits four bosonic zero modes $x^{\mu}$
corresponding to the breakdown of four-dimensional Poincar{\'e}
invariance and Goldstino modes $\theta^{\alpha}_A$ and $\ov
\tau^{\dot{\alpha}}_A$ associated with the supersymmetries broken by
the instanton.\footnote{This is the minimal content of neutral zero
  modes. If $\Sigma$ is not rigid, as in the $\cN=4$ gauge instanton,
  then we also have moduli for the broken translation
  (super)symmetries in the internal space.} So far this is just as for
the gauge instanton case discussed in
section~\ref{sec:gauge-review}. Remember that there we also had a
background space-filling brane wrapping $\Sigma$. In this section we
want to study in detail what happens when the background D6 brane is
wrapping a cycle $\Sigma'$ homologically distinct to
$\Sigma$. Generically there will be intersections between the
instanton and the brane, and thus there will be some charged open
string modes localized in there. These modes will be the focus of our
analysis.

We start in section \ref{sec:D-brane-instanton-system} by analyzing
the instanton spectrum and determining the supersymmetry
transformations in the case when the D-instanton D-brane system
preserves some common supersymmetry. This analysis relies heavily on
CFT techniques. At the end of section
\ref{sec:D-brane-instanton-system} we summarize our results. In
section \ref{sec:effective-action} we determine the effective
instanton action by calculating various string amplitudes and discuss
its supersymmetric nature. Finally in section \ref{sec:saturation} we
discuss the saturation of the instanton zero modes by computing the
path integral.

\subsection{Description of the D-brane-instanton modes}
\label{sec:D-brane-instanton-system}

The different massive and massless charged instanton modes appear as
excitations of the NS- and R-vacuum respectively. Schematically they
take the form
\begin{equation}
\sum_{k \epsilon \mathbf{Z}} \left(\alpha^I_{k-\theta_I}\right)^{N^I_k}
\left(\psi^I_{k+\frac{1}{2}-\theta_I}\right)^{M^I_k} \,
|0\rangle^{NS} \qquad \sum_{k \epsilon
\mathbf{Z}}\left(\alpha^I_{k-\theta_I}\right)^{N^I_k}
\left(\psi^I_{k-\theta_I}\right)^{M^I_k} \, |0\rangle^{R}\,\,,
\label{eq:states}
\end{equation}
where $\theta_I$ is the intersection angle between the D-brane and the
instanton in the $I$-th dimension\footnote{We take $\theta_I$ to be
  defined in the range $[-1,1)$.} and
$\alpha^I_k$ and $\psi^I_k$ denote the bosonic and fermionic creation
operators. Note that the fermionic creation operator $\psi^I_k$ has
Fermi statistics, and thus $M^I_k$ can only take the values $0$ or
$1$, while $N^I_{k}$ can be any non-negative integer.

The mass of these states is given by:
\begin{align}
\text{NS-sector:}&\qquad \qquad  M^2 \sim \varepsilon^{NS}_0 +
\sum_{k}\Big( N^I_k (k-\theta_I) + M^I_k
(k+\frac{1}{2}-\theta_I)\Big)
\label{eq:bosonic-states}
\\
\text{R-sector:}&\qquad \qquad  M^2 \sim \varepsilon^R_0 +
\sum_{k}\Big( N^I_k (k-\theta_I) + M^I_k (k-\theta_I)\Big)\,\,,
\end{align}
where $\varepsilon_0^{NS,R}$ denotes the zero point energy in the NS
or R sectors, and crucially depends on the intersection angles of the
D6-brane and the E2-instanton. Note that due to the Dirichlet-Neumann
boundary conditions in space-time\footnote{That implies that the
  intersection angles in space-time are $\theta_4=\theta_5=\frac{1}{2}$
  in equation \eqref{eq:states}. From now on we ignore any excitations
  in space-time which are genuinely stringy in the sense that there
  mass is non-zero even if the instanton and D-brane wrap the same
  cycle.} the zero point energy for the NS sector is shifted by $1/2$
compared to the D6-D6 brane system. This implies that there are no
tachyonic modes between the D-brane and the instanton and for
non-trivial angles the bosonic modes are always massive. The absence
of tachyons in this setup rules out the possibility of recombination
of the instanton and D-brane.

In this work we will focus only on instanton modes whose mass takes the form
\begin{equation}
M^2 \sim \sum^3_{I=1} a^I \theta_I \,\,.\label{eq:mass-term}
\end{equation}
thus we ignore states which are genuinely stringy in the sense that
even in the limit $\theta_I \rightarrow 0$ they are massive. That
restricts the bosonic and fermion number $N^I_k$ and $M^I_k$. For the
NS sector no fermionic excitations are allowed and only $N^I_k$ can be
non-vanishing. Thus the only states in the NS-sector are the vacuum
and a tower built up from an arbitrary number of excitations
$\alpha^I_{-\theta_I}$
\begin{align}
\label{eq:NS-states}
\sum_{I} \left(\alpha^I_{-\theta_I}  \right)^{N^I_0} \,\, | 0\rangle^{NS}        \,\,.
\end{align}
For the R-sector, the restriction to modes with masses of the type
\eqref{eq:mass-term} implies that $N^I_k=M^I_k=0$ for $k>0$. Thus the
generic state in R-sector has the form
 \begin{align}
\label{eq:R-states}
\sum_{I} \left(\alpha^I_{-\theta_I}  \right)^{N^I_0} \, \left(\psi^I_{-\theta_I}  \right)^{M^I_0} \, | 0\rangle^{R}        \,\,.
\end{align}
In both~\eqref{eq:NS-states} and \eqref{eq:R-states} we find an
infinite tower of states with similar properties, differing only on
their mass. In the following we will restrict our analysis to the
lowest states in the tower. Higher levels in the tower work in the
same way.

The states \eqref{eq:states} are subject to the GSO-projection which
puts an additional constraint on the total number of fermionic
creation operators in the R-sector. Depending on the intersection
angles the number has to be odd or even. For the CPT conjugated sector
the GSO-projection projects onto states whose total fermionic creation
operator number is even if it was odd for the original sector and vice
versa.

There is a one to one correspondence between the states and the
respective vertex operators which has been analyzed for different
intersection angles in \cite{Cvetic:2006iz} for an intersecting D6
brane setup\footnote{For string amplitude calculations involving
  states arising from intersecting D6-branes see for instance
  \cite{Cvetic:2003ch,Abel:2003vv,Klebanov:2003my,Abel:2003yx,Lust:2004cx,Bertolini:2005qh,Cvetic:2006iz}
  and references therein.  }. The difference to the E2-D6 setup we are
investigating are the Neumann-Dirichlet (ND) boundary conditions in
space-time between the instanton and the D-brane. For the vertex
operators this implies the presence of twist fields in space-time
associated with ND boundary conditions. Moreover, in the R-sector only
states whose $U(1)$-wordsheet charge is $Q_{WS}= -\frac{1}{2} \,\,
\text{mod} \,\,2$ are allowed, thus only chiral modes survive the
GSO-projection \cite{Blumenhagen:2006xt,Ibanez:2006da,Florea:2006si,
  Cvetic:2007ku}. The latter is related to the fact that the total
fermion number in the R-sector is constrained due to the
GSO-projection.

In the following we analyze the instanton mode structure for a concrete setup, we determine the instanton spectrum their corresponding vertex operators as well as the SUSY transformations, in case the D-instanton D-brane system preserves some common supersymmetry. 
Later in subsection \ref{sec:effective-action} we derive the interaction terms involving charged and neutral instanton modes. The latter are crucial for the
saturation of fermionic instanton modes in the path integral and thus
important for answering the question of whether a generic rigid $U(1)$
instanton gives contributions to the superpotential.

Without loss of generality we choose the sign of the angles to be:
\begin{equation}
\theta_1>0 \qquad \theta_2>0 \qquad \theta_3<0 \,\,,\label{eq:setup}
\end{equation}
where we require the supersymmetry condition
\begin{equation}
\theta_1+\theta_2+\theta_3=0\,\,.
\end{equation}
The preserved supercharge is ${\ov Q}^0_{\dot{\alpha}}$  while all
the others ${\ov Q}^A_{\dot{\alpha}}$ with $A=1,2,3$ are broken
unless some angles are trivial as we will see later. Let us display
the form of the supercharges explicitly since we will make use of
them extensively in the following
\begin{equation} \ov Q^A_{\dot{\alpha}}= S_{\dot{\alpha}}
\, S^A \, e^{-\varphi/2} \label{eq:supercharges}
 \end{equation}
where $S_{\dot{\alpha}}$ is the spin field in space-time with negative chirality and $S^A$ the spin fields in the internal six dimension with negative chirality. The latter can be bosonized and then take the form
\begin{equation}
S^0=\prod^3_{I=1} e^{-\frac{i}{2}H_I} \qquad S^A=
e^{\frac{i}{2}H_A}\,\prod^3_{I\neq A\neq 0} e^{-\frac{i}{2}H_I}\,\,.
\end{equation}

We will start with the states which arise from the $E2{-}a$ sector,
where $a$ denotes the D6-brane. These states transform in the $(E2,\ov
a)$ representation under the gauge groups living on the instanton and
D-brane (that is, they are bifundamentals). We will analyze first the
states arising from the NS sector. If we require the mass to take the
form \eqref{eq:mass-term} and ignore any bosonic excitation the only
state is the vacuum
\begin{equation}
|0\rangle^{NS}_{E2a}
\end{equation}
whose vertex operator takes the form
\begin{equation}
V^{-1}_{\omega_{\dot{\alpha}}}(z) =\omega_{\dot{\alpha}}\,
S^{\dot{\alpha}} \, \Sigma\, e^{i\theta_1H_1} \, \sigma_{\theta_1}\,
e^{i\theta_2H_2} \, \sigma_{\theta_2}\, e^{i\theta_3H_3} \,
\sigma_{1+\theta_3} \, e^{-\varphi}\,\,,
\label{eq:vertexop-omega}
\end{equation}
Here $\Sigma$ denotes the bosonic twist field which ensures Dirichlet-Neumann boundary conditions in space-time and $\sigma_{\theta_I}$ and $^{i\theta_IH_I}$ are the bosonic and fermionic twist fields ensuring the boundary conditions in the $I$-th complex internal dimension. The mass of $|0\rangle^{NS}_{E2a}$ is simply given by the zero-point energy in
\eqref{eq:bosonic-states} which for the this specific configuration is
$\frac{1}{2}(\theta_1+\theta_2-\theta_3)$.

For the R-sector one observes four different states surviving the GSO
projection. Note that with the conventions in \eqref{eq:setup}, the
total fermionic operator number in the $E2{-}a$ sector has to be even.
Again we are ignoring any additional bosonic excitations. These four
states are given by:
\begin{equation*}
|0\rangle^R_{E2a} \qquad \psi_{-\theta_1} \, \psi_{-\theta_2}
|0\rangle^R_{E2a} \qquad \psi_{-\theta_1} \, \psi_{\theta_3}
|0\rangle^R_{E2a}\qquad \psi_{-\theta_2} \, \psi_{\theta_3}
|0\rangle^R_{E2a}\,\,.
\end{equation*}

The masses and vertex operators (in the $-\frac{1}{2}$ ghost picture)
for these states are displayed in the following expressions:
\begin{itemize}
\item[$\bullet$] $\mu_0=\psi_{-\theta_1} \, \psi_{-\theta_2} |0\rangle^R_{E2a}$ \qquad $M^2=\theta_1+\theta_2$
\begin{align}
V^{-\frac{1}{2}}_{\mu_0}(z) =\mu_{0} \, \Sigma
e^{i(\theta_1+\frac{1}{2})H_1} \, \sigma_{\theta_1}\,
e^{i(\theta_2+\frac{1}{2})H_2} \, \sigma_{\theta_2}\,
e^{i(\theta_3+\frac{1}{2})H_3} \, \sigma_{1+\theta_3} \,
e^{-\varphi/2} \label{eq:vertexoperatormu-0 }
\end{align}
\item[$\bullet$]  $\mu_1=\psi_{-\theta_1} \, \psi_{\theta_3} |0\rangle^R_{E2a}$ \qquad $M^2=\theta_1-\theta_3$
\begin{align*}
V^{-\frac{1}{2}}_{\mu_1}(z) =\mu_1 \, \Sigma
e^{i(\theta_1+\frac{1}{2})H_1} \, \sigma_{\theta_1}\,
e^{i(\theta_2-\frac{1}{2})H_2} \, \sigma_{\theta_2}\,
e^{i(\theta_3-\frac{1}{2})H_3} \, \sigma_{1+\theta_3} \,
e^{-\varphi/2}
\end{align*}
\item[$\bullet$] $\mu_2=\psi_{-\theta_2} \, \psi_{\theta_3} |0\rangle^R_{E2a}$ \qquad $M^2=\theta_2 -\theta_3$
\begin{align*}
V^{-\frac{1}{2}}_{\mu_2}(z) =\mu_2 \, \Sigma
e^{i(\theta_1-\frac{1}{2})H_1} \, \sigma_{\theta_1}\,
e^{i(\theta_2+\frac{1}{2})H_2} \, \sigma_{\theta_2}\,
e^{i(\theta_3-\frac{1}{2})H_3} \, \sigma_{1+\theta_3} \,
e^{-\varphi/2}
\end{align*}
\item[$\bullet$] $\mu_3=|0\rangle^R_{E2a}$ \qquad $M^2=0$
\begin{align*}
V^{-\frac{1}{2}}_{\mu_3}(z) =\mu_3 \, \Sigma\,
e^{i(\theta_1-\frac{1}{2})H_1} \, \sigma_{\theta_1}\,
e^{i(\theta_2-\frac{1}{2})H_2} \, \sigma_{\theta_2}\,
e^{i(\theta_3+\frac{1}{2})H_3} \, \sigma_{1+\theta_3} \,
e^{-\varphi/2}
\end{align*}
\end{itemize}
In the existing literature the massless charged mode $\mu_3$ is often
denoted by $\lambda$.

In the sequel we will apply the four supercharges
\eqref{eq:supercharges} to the fermionic fields. We will derive under
which circumstances the supercharges are preserved and the
transformation behavior of the fermionic fields under the respective
supercharge. In order to do this we need to know the explicit form of
the OPE's of various conformal fields, they are given by the following
expressions:
\begin{equation}
\label{eq:OPEs}
\begin{split}
e^{k\, \varphi(z)} e^{l\, \varphi(w)} &\sim(z-w)^{-k\, l} e^{(k+l)\varphi(z)}\\
S^{\dot{\alpha}}(z)\, S^{\dot{\beta}}(w)&\sim \epsilon^{\dot{\alpha}\, \dot{\beta}}(z-w)^{-\frac{1}{2}}\\ e^{i k H_I (z)} \, e^{i l H_{J}(w)} &\sim
\delta^{IJ} (z-w)^{-k \, l} e^{i(k+l) H_{I}(z)}
\end{split}
\end{equation}

With this information is it simple to obtain the supersymmetry
transformations of the conformal fields, they are computed as follows:
\begin{align*}
[\ov \xi^{\dot{\alpha}}_0\, \ov Q^0_{\dot{\alpha}},
V^{-\frac{1}{2}}_{\mu_0}]& =\xi^{\dot{\alpha}}_0 \cdot \mu_0  \oint
dw e^{-\varphi/2(w)}\, e^{-\varphi/2(z)} S_{\dot{\alpha}}(w)
\Sigma(z) e^{-\frac{i}{2}H_1(w)}e^{i(\theta_1+\frac{1}{2})H_1(z)} \\
& \qquad e^{-\frac{i}{2}H_2(w)} e^{i(\theta_2+\frac{1}{2})H_2(z)}
e^{-\frac{i}{2}H_3(w)} e^{i(\theta_3+\frac{1}{2})H_3(z)}
 \sigma_{\theta_1}(z)\,
\sigma_{\theta_2}(z)\, \sigma_{1+\theta_3}(z)\\&=
\xi^{\dot{\alpha}}_0 \cdot \mu_0  \oint dw \frac{
e^{-\varphi(z)}\,S^{\dot \alpha}\, \Sigma e^{-\varphi}\,
e^{i\theta_1H_1}\, \sigma_{\theta_1}\,
e^{i\theta_2H_2}\,\sigma_{\theta_2}e^{i\theta_3H_3}\,
\sigma_{1+\theta_3}}{(z-w)^{1+\frac{1}{2}(\theta_1
+\theta_2+\theta_3)}}\,\,.
\end{align*}
Thus for $\theta_1+\theta_2+\theta_3=0$ we recover the SUSY
transformation
\begin{align}
[\ov \xi^{\dot{\alpha}}_0\, \ov Q^0_{\dot{\alpha}},
V^{-\frac{1}{2}}_{\mu_0}]= \delta_{\ov \xi_0}\,
V^{-1}_{\omega_{\dot{\alpha}}}\,\,.
\label{eq:susytransf1a}
\end{align}
Similarly we obtain for the other fermionic states
\begin{align} \nonumber
-\theta_1+\theta_2+\theta_3&=0\,:\qquad [\ov \xi^{\dot{\alpha}}_1\,
\ov Q^1_{\dot{\alpha}}, V^{-\frac{1}{2}}_{\mu_1}]= \delta_{\ov
\xi_1}\, V^{-1}_{\omega_{\dot{\alpha}}}
\\\,\,
\theta_1-\theta_2+\theta_3&=0\,: \qquad [\ov \xi^{\dot{\alpha}}_2\,
\ov Q^2_{\dot{\alpha}}, V^{-\frac{1}{2}}_{\mu_2}]= \delta_{\ov
\xi_2}\, V^{-1}_{\omega_{\dot{\alpha}}}
\label{eq:susytransf1b}
\\
\theta_1+\theta_2-\theta_3&=0\,: \qquad [\ov \xi^{\dot{\alpha}}_3\,
\ov Q^3_{\dot{\alpha}}, V^{-\frac{1}{2}}_{\mu^3}]= \delta_{\ov
\xi_3}\, V^{-1}_{\omega_{\dot{\alpha}}}\,\,.
\nonumber
\end{align}

Applying the same supercharge again we will get additional fermionic
states. As we will see momentarily the presence of these states is
required to define a proper mass term. Again we will be explicit for
$\ov Q^0_{\dot{\alpha}}$ and state the result for the other
supercharges,
\begin{align*}
[\ov \xi^{\dot{\alpha}}_0\, \ov Q^0_{\dot{\alpha}}(w),
V^{-1}_{\omega}(z)] =&\, \ov \xi_0 \cdot \omega \oint dw
\frac{e^{-3\varphi/2}\, \Sigma\, e^{i(\theta_1-\frac{1}{2})H_1}\,
\,\sigma_{\theta_1} \,e^{i(\theta_2-\frac{1}{2})H_2} \,
\sigma_{\theta_2}\,
e^{i(\theta_3-\frac{1}{2})H_3}\,\sigma_{1+\theta_3}}{(z-w)^{1+\frac{1}{2}(\theta_1+\theta_2+\theta_3)}}\,\,.
\end{align*}
Here we applied the OPE's given in \eqref{eq:OPEs} and we recovered,
as expected, that $\ov Q^0_{\dot{\alpha}}$ is preserved for
$\theta_1+\theta_2+\theta_3=0$. The SUSY transformation behavior is
\begin{equation}
[\xi^{\dot{\alpha}}_0 \, \ov Q^0_{\dot{\alpha}},
V^{-1}_{\omega}]=\delta_{\xi^0}\, V^{-\frac{3}{2}}_{\mu'_0}\,\,.
\label{eq:susytransf2a}
\end{equation}
The vertex operator of the associated state $\mu'_0$ in the
$-\frac{3}{2}$ ghost picture takes the form
\begin{align*}
V^{-\frac{3}{2}}_{\mu'_0}(z)=  \mu'_0\, e^{-3\varphi/2}\, \Sigma\,
e^{i(\theta_1-\frac{1}{2})H_1}\, \,\sigma_{\theta_1}
\,e^{i(\theta_2-\frac{1}{2})H_2} \, \sigma_{\theta_2}\,
e^{i(\theta_3-\frac{1}{2})H_3}\,\sigma_{1+\theta_3}\,\,.
\end{align*}
The conformal dimension of $V^{-\frac{3}{2}}_{\mu'_0}$ is $1-\theta_3$
and thus the mass is expected to be $M^2=-\theta_3$.  Let us apply the
picture changing operator to get the vertex operator in the canonical
$-\frac{1}{2}$ ghost picture. After applying the picture changing
procedure
\begin{equation}
\lim_{w\rightarrow z} \, O_{PCO}(w) \, V^{-\frac{3}{2}}_{\mu_0}(z)=
V^{-\frac{1}{2}}_{\mu_0}(z)\,\,,
\end{equation}
we obtain the vertex operator in the canonical $-\frac{1}{2}$-ghost
picture, which allows us to identify the corresponding state. The
picture changing operator is given by
\begin{align}
O_{PCO}(z)=e^{\varphi(z)} T_F(z)\,\,,
\end{align}
where $T_F$ is
\begin{align}
T_F(z)=\psi^{\mu}(z) \, \partial X^{\mu}(z) + e^{i H_I(z)}\,
\partial {\ov Z}^I +  e^{-i H_I(z)}\, \partial Z^{I}\,\,,
\end{align}
where $\partial Z^I$ and $e^{i H_I(z)}$ are the complexified bosonic and fermionic fields in the $I$-th internal dimensions, 
whereas $\partial X^{\mu}$ and $\psi^{\mu}$ denote the  bosonic and fermionic fields in space-time.
Using the OPE's displayed in \eqref{eq:OPEs} and below \cite{Dixon:1986qv}\footnote{The OPE's with space-time fields do not play any crucial role here.}
\begin{equation}
\label{eq:OPE's-twist}
\begin{split}
\partial Z(z) \sigma_{\theta}(w) &\sim (z-w)^{\theta-1} \,
\tau_{\theta}(z)\\
\partial {\ov Z}(z) \sigma_{\theta}(w) &\sim (z-w)^{-\theta} \,
\tilde{\tau}_{\theta}(z)\\
\partial Z(z) \sigma_{1-\theta}(w) &\sim (z-w)^{-\theta} \,
\tilde{\tau}_{1-\theta}(z)\\
\partial {\ov Z}(z) \sigma_{1-\theta}(w) &\sim (z-w)^{\theta-1} \,
\tau_{1-\theta}(z)
\end{split}
\end{equation}
the vertex
operator for $\mu'_0$ in the $-\frac{1}{2}$-ghost picture takes the
form
\begin{align*}
V^{-\frac{1}{2}}_{\mu'_0}(z) =\mu'_{0} \, \Sigma
e^{i(\theta_1-\frac{1}{2})H_1} \, \sigma_{\theta_1}\,
e^{i(\theta_2-\frac{1}{2})H_2} \, \sigma_{\theta_2}\,
e^{i(\theta_3+\frac{1}{2})H_3} \, \tau_{1+\theta_3} \,
e^{-\varphi/2}\,\,.
\end{align*}
We read off that the corresponding state is given by
$\mu'_0=\alpha_{\theta_3} \ket{0}^R_{E2a}$, which indeed has the right
mass $M^2=-\theta_3$. Analogously we obtain the other three states
$\mu'_1$, $\mu'_2$ and $\mu'_3$
\begin{itemize}
\item[$\bullet$]  $\mu'_1=\alpha_{-\theta_2} |0\rangle^R_{E2a}$ \qquad $M^2=\theta_2$
\begin{align*}
V^{-\frac{1}{2}}_{\mu'_1}(z) =\mu'_1 \, \Sigma
e^{i(\theta_1-\frac{1}{2})H_1} \, \sigma_{\theta_1}\,
e^{i(\theta_2-\frac{1}{2})H_2} \, \tau_{\theta_2}\,
e^{i(\theta_3+\frac{1}{2})H_3} \, \sigma_{1+\theta_3} \,
e^{-\varphi/2}
\end{align*}
\item[$\bullet$]  $\mu'_2=\alpha_{-\theta_1} |0\rangle^R_{E2a}$ \qquad $M^2=\theta_1$
\begin{align*}
V^{-\frac{1}{2}}_{\mu'_2}(z) =\mu'_2 \, \Sigma
e^{i(\theta_1-\frac{1}{2})H_1} \, \tau_{\theta_1}\,
e^{i(\theta_2-\frac{1}{2})H_2} \, \sigma_{\theta_2}\,
e^{i(\theta_3+\frac{1}{2})H_3} \, \sigma_{1+\theta_3} \,
e^{-\varphi/2}
\end{align*}
\item[$\bullet$] $\mu'_3=\frac{1}{3}\left[\psi_{-\theta_2}\,\psi_{\theta_3}\,
\alpha_{-\theta_1}+\psi_{-\theta_1}\,\psi_{\theta_3}\,
\alpha_{-\theta_2}+\psi_{-\theta_1}\,\psi_{-\theta_2}\,
\alpha_{\theta_3}\right]|0\rangle^R_{E2a}$\qquad
$M^2=\theta_1+\theta_2-\theta_3$
\begin{align*}
V^{-\frac{1}{2}}_{\mu'_3}(z)=  \mu'_3\, e^{-\varphi/2}\, \Sigma\, &
\left[ e^{i(\theta_1-\frac{1}{2})H_1}\, \tau_{\theta_1}
\,e^{i(\theta_2+\frac{1}{2})H_2} \, \sigma_{\theta_2}\,
e^{i(\theta_3-\frac{1}{2})H_3}\,\sigma_{1+\theta_3} \right.
\\&\left.+
e^{i(\theta_1+\frac{1}{2})H_1}\, \,\sigma_{\theta_1}
\,e^{i(\theta_2-\frac{1}{2})H_2} \, \tau_{\theta_2}\,
e^{i(\theta_3-\frac{1}{2})H_3}\,\sigma_{1+\theta_3}\right.\\
& \left.+ e^{i(\theta_1+\frac{1}{2})H_1}\, \,\sigma_{\theta_1}
\,e^{i(\theta_2+\frac{1}{2})H_2} \, \sigma_{\theta_2}\,
e^{i(\theta_3+\frac{1}{2})H_3}\,\tau_{1+\theta_3}\right]\,\,.
\end{align*}
\end{itemize}

Moreover, analogously to  before one can show that $\mu'_A$ is related to $\omega$ if the supercharge $\ov Q^A_{\dot{\alpha}}$ is preserved
\begin{align} \nonumber
[\xi^{\dot{\alpha}}_1 \, \ov Q^1_{\dot{\alpha}},
V^{-1}_{\omega}]&=\delta_{\xi^1}\, V^{-\frac{3}{2}}_{\mu'_1}\\
[\xi^{\dot{\alpha}}_2 \, \ov Q^2_{\dot{\alpha}},
V^{-1}_{\omega}]&=\delta_{\xi^2}\, V^{-\frac{3}{2}}_{\mu'_2}
\label{eq:susytransf2b}
\\
[\xi^{\dot{\alpha}}_3 \, \ov Q^3_{\dot{\alpha}},
V^{-1}_{\omega}]&=\delta_{\xi^3}\, V^{-\frac{3}{2}}_{\mu'_3}\,\,.
\nonumber
\end{align} 

From \eqref{eq:susytransf1a}, \eqref{eq:susytransf1b}, \eqref{eq:susytransf2a} and \eqref{eq:susytransf2b} we can read off the SUSY transformations of the instanton modes $\mu_A$, $\omega$ and $\mu'_A$
\begin{align}
\delta_{\ov \xi^{A}} \mu'_A = i\,\ov \xi^A_{\dot{\alpha}} \omega^{\dot{\alpha}} \,\,\,\,\,\,\,\,\,\,\,\,\,\,\delta_{\ov
\xi^{A}} \omega_{\dot{\alpha}}= i \ov \xi^A_{\dot{\alpha}} \mu_A\,\,.
\end{align}

Let us turn to the CPT-conjugated sector $a{-}E2$. Again ignoring any
bosonic excitations there is only one state in the NS sector which is
simply given by the vacuum $|0\rangle^{NS}_{aE2}$. Its mass is
$M^2=\frac{1}{2}(\theta_1+\theta_2-\theta_3)$ and the vertex operator
takes the form
\begin{equation}
V^{-1}_{\ov \omega_{\dot{\alpha}}}(z) =\ov \omega_{\dot{\alpha}}\,
S^{\dot{\alpha}} \, \Sigma\, e^{-i\theta_1H_1} \,
\sigma_{1-\theta_1}\, e^{-i\theta_2H_2} \, \sigma_{1-\theta_2}\,\,
e^{-i\theta_3H_3} \, \sigma_{-\theta_3} \, e^{-\varphi}\,\,.
\label{eq:vertexoperatoromegabar}
\end{equation}
Note that $\omega$ and $\ov \omega$ have the same mass and one can
show the presence of the mass term
\begin{equation}
M_{\omega} \omega \, \ov \omega
\end{equation}
by computing
\begin{equation}
\langle V^{-1}_{\omega}(z_1) \,   V^{-1}_{\ov \omega}(z_2)
\rangle\,\,.
\end{equation}
For the R-sector the GSO-projection projects onto states with odd
total fermion number. Thus neglecting bosonic excitations the R-sector
exhibits four different states
\begin{align}
\psi_{-\theta_1}|0\rangle^{R}_{aE2} \qquad
\psi_{-\theta_2}|0\rangle^{R}_{aE2}\qquad
\psi_{\theta_3}|0\rangle^{R}_{aE2} \qquad \psi_{-\theta_1}\,
\psi_{-\theta_2}\, \psi_{\theta_3}|0\rangle^{R}_{aE2}\,\,.
\end{align}
Below we display the corresponding vertex operators in the
$-\frac{1}{2}$-ghost picture.
\begin{itemize}
\item[$\bullet$] $\ov \mu_0=\psi_{\theta_3}\ket{0}^{R}_{aE2} $\qquad $M^2=-\theta_3$\\
\begin{align}
V^{-\frac{1}{2}}_{{\ov \mu_0}}(z)= {\ov \mu_0} \, \Sigma
e^{-i(\theta_1-\frac{1}{2})H_1} \, \sigma_{1-\theta_1}\,
e^{-i(\theta_2-\frac{1}{2})H_2} \, \sigma_{1-\theta_2}\,
e^{-i(\theta_3-\frac{1}{2})H_3} \, \sigma_{-\theta_3} \,
e^{-\varphi/2}
\label{eq:vertexop-mu-0}
\end{align}
\item[$\bullet$] $\ov \mu_1=\psi_{-\theta_2}|0\rangle^{R}_{aE2} $\qquad $M^2=\theta_2$\\
\begin{align*}
V^{-\frac{1}{2}}_{\ov \mu_1} (z)=  \ov \mu_{1} \, \Sigma
e^{-i(\theta_1-\frac{1}{2})H_1} \, \sigma_{1-\theta_1}\,
e^{-i(\theta_2+\frac{1}{2})H_2} \, \sigma_{1-\theta_2}\,
e^{-i(\theta_3+\frac{1}{2})H_3} \, \sigma_{-\theta_3} \,
e^{-\varphi/2}
\end{align*}
\item[$\bullet$] $\ov \mu_2=\psi_{-\theta_1}|0\rangle^{R}_{aE2} $\qquad $M^2=\theta_1$\\
\begin{align*}
V^{-\frac{1}{2}}_{\ov \mu_2} (z)=  \ov \mu_{2} \, \Sigma
e^{-i(\theta_1+\frac{1}{2})H_1} \, \sigma_{1-\theta_1}\,
e^{-i(\theta_2-\frac{1}{2})H_2} \, \sigma_{1-\theta_2}\,
e^{-i(\theta_3+\frac{1}{2})H_3} \, \sigma_{-\theta_3} \,
e^{-\varphi/2}
\end{align*}
\item[$\bullet$] $\ov \mu_3=\psi_{-\theta_1}\,\psi_{-\theta_2}\,\psi_{\theta_3}|0\rangle^{R}_{aE2} $\qquad $M^2=\theta_1+\theta_2-\theta_3$\\
\begin{align*}
V^{-\frac{1}{2}}_{{\ov \mu_3}}(z)= {\ov \mu_3} \, \Sigma
e^{-i(\theta_1+\frac{1}{2})H_1} \, \sigma_{1-\theta_1}\,
e^{-i(\theta_2+\frac{1}{2})H_2} \, \sigma_{1-\theta_2}\,
e^{-i(\theta_3-\frac{1}{2})H_3} \, \sigma_{-\theta_3} \,
e^{-\varphi/2}
\end{align*}
\end{itemize}
In the case where all supersymmetry is broken the mass of $\ov \mu_A$
is different from the mass of $\mu_A$, and thus the mass terms for the
fermionic modes cannot be of the form $\mu_A \ov \mu_A$ but rather are
given by
\begin{equation}
\label{mass term 1}
\mu'_0\, \ov \mu_0 \qquad  \mu'_1\, \ov \mu_1 \qquad  \mu'_2\, \ov
\mu_2\qquad  \mu'_3\, \ov \mu_3\,\,.
\end{equation}
Again this can be verified by computing the string disc amplitude
\begin{equation}
\vev{V^{-\frac{3}{2}}_{\mu'_A} (z_1) \, V^{-\frac{1}{2}}_{\ov \mu_A}
(z_2)}\,\,.
\end{equation}
In order to ensure ghost charge -2 of the whole amplitude we need the
vertex operator of $\mu'_A$ in the $-\frac{3}{2}$ ghost picture which
ensures the conservation of $U(1)$ world sheet charge.

Similarly there should be three $\ov \mu'$ states which give together
with $\mu_0$, $\mu_1$ and $\mu_2$ the respective mass terms. Indeed
acting with the supercharges $\ov Q^A_{\dot{\alpha}}$ with $A=0,1,2$
on the bosonic state $\ov \omega_{\dot{\alpha}}$, analogously to the
procedure performed for the sector $E2{-}a$, we obtain the states
\begin{itemize}
\item[$\bullet$] $\ov
\mu'_0=\frac{1}{2}\left(\psi_{-\theta_1} \,\alpha_{-\theta_2}+
\psi_{-\theta_2} \,\alpha_{-\theta_1}\right) |0\rangle^R_{aE2}$ \qquad $M^2=\theta_1+\theta_2$\\
\begin{align*}
V^{-\frac{1}{2}}_{{\ov \mu}'}(z)&= {\ov \mu'_{0}} \, \Sigma
e^{-i(\theta_3-\frac{1}{2})H_3} \, \sigma_{-\theta_3} \,
e^{-\varphi/2}\\ &\times \left[e^{-i(\theta_1-\frac{1}{2})H_1} \,
\tau_{1-\theta_1} e^{-i(\theta_2+\frac{1}{2})H_2} \,
\sigma_{1-\theta_2}+ e^{-i(\theta_1+\frac{1}{2})H_1} \,
\sigma_{1-\theta_1} e^{-i(\theta_2-\frac{1}{2})H_2} \,
\tau_{1-\theta_2}\right]
\end{align*}
\item[$\bullet$] $\ov \mu'_1=\psi_{\theta_3} \, \alpha_{-\theta_1}|0\rangle^R_{aE2} $\qquad $M^2=\theta_1-\theta_3$\\
\begin{align*}
V^{-\frac{1}{2}}_{\ov \mu'_1}(z) =\ov \mu'_1\,
 \Sigma\, e^{-i(\theta_1+\frac{1}{2}) H_1} \,
\tau_{1-\theta_1}\, e^{-i(\theta_2+\frac{1}{2})H_2} \,
\sigma_{1-\theta_2}\, e^{-i(\theta_3-\frac{1}{2})H_3} \,
\sigma_{-\theta_3} \, e^{-\varphi/2}
\end{align*}
\item[$\bullet$] $\ov \mu'_2=\psi_{\theta_3} \,
\alpha_{-\theta_2}|0\rangle^R_{aE2}$\qquad $M^2=\theta_2-\theta_3$\\
\begin{align*}
V^{-\frac{1}{2}}_{\ov \mu'_2}(z) =\ov \mu'_2\,
 \Sigma\, e^{-i(\theta_1-\frac{1}{2}) H_1} \,
\sigma_{1-\theta_1}\, e^{-i(\theta_2-\frac{1}{2})H_2} \,
\tau_{1-\theta_2}\, e^{-i(\theta_3-\frac{1}{2})H_3} \,
\sigma_{-\theta_3} \, e^{-\varphi/2}
\end{align*}
\end{itemize}
Again one can show by computing the two point amplitude
\begin{equation}
\vev{V^{-\frac{1}{2}}_{\mu_A} (z_1) \, V^{-\frac{3}{2}}_{\ov \mu'_A}
(z_2)}
\end{equation}
that the mass terms
\begin{equation}
\mu_0\, \ov \mu'_0 \qquad  \mu_1\, \ov \mu'_1 \qquad  \mu_2\, \ov
\mu'_2
\label{mass term 2}
\end{equation}
are indeed present. Note that there is no state $\ov \mu'_3$. This is
expected since $\mu_3$ is massless, and hence no mass term of the form
$\mu_3 \, \ov \mu'_3$ should appear in the instanton
action\footnote{Recall that $\mu_3$ is often refered to as $\lambda$
  mode in the existing literature.}.

Analogously to the $E2{-}a$ sector, in case the D-instanton and the
D-brane preserve some common supersymmetry one can show by applying
the supercharges $\ov Q^A_{\dot {\alpha}}$ onto the vertex operators
of $\ov \mu_A$, $\ov \omega$ and $\ov \mu'_A$ that the instanton states
are related to each other via the following transformations
\begin{align}
\delta_{\ov \xi^{A}} \ov\mu'_A = i\,\ov \xi^A_{\dot{\alpha}} \ov\omega^{\dot{\alpha}} \,\,\,\,\,\,\,\,\,\,\,\,\,\, \delta_{\ov
    \xi^{A}} \ov\omega_{\dot{\alpha}}= i \ov \xi^A_{\dot{\alpha}} \ov\mu_A\,\,.
\end{align}
\subsubsection*{Brief summary on the spectrum and SUSY transformation }
\vspace{3mm}

The D-instanton D-brane configuration considered above  exhibits the intersection pattern
\begin{equation}
\theta_1>0 \qquad \theta_2>0 \qquad \theta_3<0 \,\,.
\end{equation}
We determined for such a configuration all states whose mass squared is at most of linear order
in the intersecting angles $\theta_I$. In table \ref{table spectrum
  instanton} we present all states and their corresponding mass.
  
\begin{table}[h] \centering
\begin{tabular}{|c|c|c|c||c|}
  \hline
  E2-a & CFT state & a-E2 & CFT state & mass$^2$\\
  \hline
  \hline
  $\mu_0$     & $\psi_{-\theta_1}\, \psi_{-\theta_2} | 0
  \rangle^R_{E2a} $  & $\ov \mu'_0$ & $\frac{1}{2} \left( 
  \psi_{-\theta_1} \alpha_{-\theta_2} + \psi_{-\theta_2} \alpha_{-\theta_1}\right)| 0 \rangle^R_{aE2} $ & $\theta_1+ \theta_2$\\
  \hline
  $\mu_1$       &$\psi_{-\theta_1}\, \psi_{\theta_3} | 0
  \rangle^R_{E2a} $  & $\ov \mu'_1$ & $\psi_{\theta_3}
  \alpha_{-\theta_1} | 0 \rangle^R_{aE2}$ & $\theta_1 -\theta_3$ \\
  \hline
  $\mu_2$      &$\psi_{-\theta_2}\, \psi_{\theta_3} | 0 \rangle^R_{E2a}$ & $\ov \mu'_2$  & $\psi_{\theta_3} \alpha_{-\theta_2} | 0 \rangle^R_{aE2}$ &  $\theta_2 -\theta_3$\\
  \hline
  $\mu_3$         & $| 0 \rangle^R_{E2a}$ & & {\bf None} & 0\\
  \hline
  $\mu'_0$         &   $\alpha_{\theta_3} | 0 \rangle^R_{E2a} $ & $\ov \mu_0$& $\psi_{\theta_3} | 0 \rangle^R_{aE2}$ & $-\theta_3 $\\
  \hline
  $\mu'_1$        & $\alpha_{-\theta_2} | 0 \rangle^R_{E2a} $ & $\ov
  \mu_1$ &$\psi_{-\theta_2} | 0 \rangle^R_{aE2}$ & $\theta_2$\\
  \hline
  $\mu'_2$             & $\alpha_{-\theta_1} | 0 \rangle^R_{E2a}$ &
  $\ov \mu_2$ & $\psi_{-\theta_1} | 0 \rangle^R_{aE2}$ & $\theta_1$\\
  \hline
  $\mu'_3$ & $\frac{1}{3} \, \epsilon^{2}_{ijk} \psi_{-\theta_i}
  \psi_{-\theta_j} \alpha_{-\theta_k}   | 0 \rangle^R_{E2a} $  & $\ov
  \mu_3$  &$\psi_{-\theta_1} \psi_{-\theta_2} \psi_{\theta_3} | 0
  \rangle^R_{aE2}$ & $\theta_1 +\theta_2 -\theta_3 $\\
  \hline
  $\omega^{\dot{\alpha}}$ & $| 0 \rangle^{NS}_{E2a}$ & $\ov
  \omega^{\dot{\alpha}}$ & $| 0 \rangle^{NS}_{E2a}$ & $\frac{1}{2}\left(\theta_1 +\theta_2 -\theta_3\right) $\\
  \hline
\end{tabular}

\caption{Instanton D-brane spectrum. We have paired modes by
  mass. Notice that, despite what the notation might suggest, $\mu$
  modes can only pair up with $\ov\mu'$ modes, and
  $\mu'$ modes can only pair up with $\ov\mu$ modes. Notice also the
  chiral spectrum at the massless level, encoded in the fact that
  $\mu_3$ has no mass partner.}
\label{table spectrum instanton}
\end{table}  

To each state corresponds a vertex operator and we have explicitly
shown the existence of the mass terms\footnote{Here $A$ runs from $0$ to $3$, while $B$ runs only from $0$ to $2$. }
\begin{equation} 
S^{mass}_{E2} = m_{\mu^A} \,\mu^A\, \ov \mu'_A + m_{\omega}\,\omega \ov \omega + m_{\ov \mu^B} \, \ov \mu^B \,\mu'_B\\,\,.
\label{eq:coupling-mass}
\end{equation}
This agrees with the result displayed in table~\ref{table
  spectrum instanton}. Note in particular that in the case where all
SUSY is broken by the E2-D6 system the masses of $\mu_A$ and $\ov
\mu_A$ are different, and thus a mass term $\mu_A \, \ov \mu^A$ is
never allowed.

 As shown above if the D-instanton D-brane system preserves the supercharge $\ov
Q^A_{\dot{\alpha}}$ the charged instanton modes are related to each
other in the following way 
\begin{align}
\delta_{\ov \xi^{A}} \mu'_A = i\,\ov \xi^A_{\dot{\alpha}} \omega^{\dot{\alpha}} \,\,\,\,\,\,\,\,\,\,\,\,\,\,\delta_{\ov
\xi^{A}} \omega_{\dot{\alpha}}= i \ov \xi^A_{\dot{\alpha}} \mu_A \,\,\,\,\,\,\,\,\,\,\,\,\,\,\delta_{\ov \xi^{A}} \ov\mu'_A = i\,\ov \xi^A_{\dot{\alpha}} \ov\omega^{\dot{\alpha}} \,\,\,\,\,\,\,\,\,\,\,\,\,\, \delta_{\ov
    \xi^{A}} \ov\omega_{\dot{\alpha}}= i \ov \xi^A_{\dot{\alpha}} \ov\mu_A\,\,.
    \label{eq: Susy transfo}
\end{align}
If $\mu_A$ or $\ov \mu_A$ is massless we expect the absence of $\ov
\mu'_A$ or $\ov \mu'_A$, respectively.  This is analogous to the D6-D6 brane configuration, where we
have in the massive case a hypermultiplet (consisting of two chiral
multiplets) while in the massless case we only have a single chiral
multiplet. In case the D-instanton and D-brane wrap the same cycle all
supercharges $\ov Q^A_{\dot\alpha}$ are preserved and all $\mu_A$ and
$\ov \mu_A$ are massless. Then the SUSY transformations reduce to the
second and fourth equations of \eqref{eq: Susy transfo} which are
exactly the usual SUSY transformations appearing in the ADHM case
\cite{Green:2000ke,Billo:2002hm}\footnote{Note that for the gauge
  instanton setup, which corresponds to all three intersection angles
  $\theta_I$ being $0$, the vacuum is defined differently. In that
  case the $\mu'_A$ and $\ov \mu'_A$ modes are projected out and do
  not appear in the instanton mode spectrum.}.

\subsection{The instanton effective action}
\label{sec:effective-action}

Above we saw that for D-branes and instantons intersecting at
non-trivial angles we get the expected mass terms for the respective
fields. Apart from these mass terms we have additional interaction
terms. Let us assume for simplicity that the instanton wraps a rigid
supersymmetric cycle on the Calabi-Yau threefold of $SU(3)$
holonomy. In this case the only neutral instanton zero modes are the
four universal bosonic zero modes $x^\mu$ and four fermionic zero
modes $\theta^{\alpha}\equiv \theta^\alpha_0$ and $\ov
\tau_{\dot{\alpha}}\equiv \ov\tau_{\dot\alpha}^0$, the rest of the
$\ov\tau^A,\theta_A$ modes being lifted by the holonomy of the
background.

In this case we would like to argue (as was also done in
\cite{Heckman:2008es}) that the ADHM-like coupling
\begin{align}
\ov \tau_{\dot{\alpha}} (\ov \omega^{\dot{\alpha}} \mu_0 + \ov \mu_0
\omega^{\dot{\alpha}})
\label{eq:coupling-tau}
\end{align}
survives even for non-trivial intersections. Indeed it is easy to show
that the $U(1)$ world sheet charge is conserved using the vertex
operators given in \eqref{eq:vertexop-omega},
\eqref{eq:vertexoperatormu-0 }, \eqref{eq:vertexoperatoromegabar},
\eqref{eq:vertexop-mu-0} and the vertex operator for $\ov \tau$
\begin{align}
V^{-\frac{1}{2}}_{\ov \tau}(z) =\ov \tau_{\dot{\alpha}} \,
S^{\dot{\alpha}} e^{-\frac{i}{2}H_1} \,e^{-\frac{i}{2}H_2} \,
e^{-\frac{i}{2}H_3}  \, e^{-\varphi/2}\,\,.
\end{align}
Note that due to the absence of the $\ov \tau^{A}$ ($A \neq 0$) the
other fermionic modes $\mu_A$ and $\ov \mu_A$ do not couple to the
bosonic modes $\ov \omega$ and $\omega$.

One might wonder if $\ov \tau$ also couples in a similar fashion to
the $\mu'_0$ and $\ov \mu'_0$ modes.  The corresponding amplitude
vanishes since it violates $U(1)$ world-sheet charge, and thus the
corresponding coupling in the instanton effective action is absent. In
fact, apart from the mass terms in \eqref{mass term 1} and \eqref{mass
  term 2}, there is no additional coupling involving $\mu'_A$ and $\ov
\mu'_A$.

There is also the coupling of the bosonic fields $\omega$ and $\ov
\omega$ to the auxiliary field $D^{\mu\nu}=\vec{D} \,
\vec{\sigma}_{\mu\nu}$
\begin{equation}
i \, \vec{D}\,\vec{W}
 \end{equation}
 where $W^c$ is displayed in \eqref{eq:Wc}. The vertex operator in the
 $0$-ghost picture of the auxiliary field $D^c$ is given by (for
 details see \cite{Billo:2002hm})
\begin{equation}
V^0_{D}=\frac{1}{2} \, \vec{D} \, \vec{ \sigma}_{\mu\nu}\, \psi^{\mu} \,
\psi^{\nu}
\end{equation}
and again it is easy to check that the $U(1)$ world sheet charge is
preserved in this amplitude.

\medskip

Summarizing, we have the following $E2{-}D6$ interaction terms:
\begin{align}
S^{E2-D6}=   m_{\ov \mu^A}\, \ov \mu^{A} \, \mu'_{A} + m_{\mu^B}
\,\mu^{B} \, \ov \mu'_{B} + m_{\omega} \,  \omega \, \ov \omega +
\ov \tau\, (\ov \omega\, \mu^0 + \ov \mu^0 \, \omega) + i
\vec{D}\cdot
  \omega \,\vec{\sigma}\,\ov\omega
  \label{eq:instD-brane-action}
\end{align}

As expected this action preserves the supercharge $\ov
Q^{0}_{\dot{\alpha}}$. This can be easily verified using the
transformation behavior for the charged and neutral instanton
displayed below (see sections \ref{sec:gauge-review} and
\ref{sec:D-brane-instanton-system}):
\begin{equation}
\label{eq:susy-transform}
\delta_{\ov \xi^{0}} \mu'_0 = i\,\ov \xi^0_{\dot{\alpha}} \omega^{\dot{\alpha}} \qquad \delta_{\ov
\xi^{0}} \omega_{\dot{\alpha}}= i \ov \xi^0_{\dot{\alpha}} \mu_0 \qquad \delta_{\ov \xi^0} \ov \tau
= \ov \xi^0\vec{\sigma} \vec{D}
\end{equation}
and similarly for the conjugated modes:
\begin{equation}
  \delta_{\ov \xi^{0}} \ov\mu'_0 = i\,\ov \xi^0_{\dot{\alpha}} \ov\omega^{\dot{\alpha}} \qquad \delta_{\ov
    \xi^{0}} \ov\omega_{\dot{\alpha}}= i \ov \xi^0_{\dot{\alpha}} \ov\mu_0
\end{equation}
For all other fields in \eqref{eq:instD-brane-action} the
transformation behavior is trivial.  Note that the presence of the
mass term for the bosons $ m_{\omega} \,\omega \ov \omega$ requires
the appearance of the mass terms $m_{\mu_0}\mu_0 \ov \mu'_0$ and
$m_{\ov \mu_0} \ov \mu_0 \mu'_0$ with $m_{\omega}=m_{\mu_0}=m_{\ov
  \mu_0} $ to ensure invariance under the supercharge $\ov
Q^{0}_{\dot{\alpha}}$. Supersymmetry also explains the lack of
couplings between $\ov\tau$ and the primed modes
$\mu'_0,\ov\mu'_0$. From~\eqref{eq:instD-brane-action} and
\eqref{eq:susy-transform} we have that in order to preserve
supersymmetry the cubic coupling to $\ov\tau$ must be of the schematic
form:
\begin{equation}
  \ov\tau\delta_{\ov\xi^0}(\omega\ov\omega)
\end{equation}
and in particular $\ov\tau$ only couples to modes that can be obtained
from supersymmetry variations of $\ov\omega$ and $\ov\omega$, which
excludes couplings to $\mu'_0$ or $\ov\mu'_0$.

\subsection{Saturation of the fermionic modes
\label{sec:saturation} }

With the explicit action \eqref{eq:instD-brane-action} for instanton
modes at hand it is a simple matter to argue that no superpotential is
generated. In order to show this, we will perform the relevant part of
the path integral calculation explicitly, and argue that it is not
possible to saturate all the integrals over fermionic modes of the
instanton simultaneously.

The path integral takes the form (here we focus only on the terms
potentially relevant for the saturation of the $\ov \tau$ modes):
\begin{equation}
\int d^4 x d^2\theta d^2 \ov \tau \prod^3_{A=0} \, d^2 \omega d^2\ov
\omega\, d  \mu'_A\, d \mu_A \prod^2_{B=0} d  \ov \mu'_B \, d\ov
\mu_B \,d \ov \mu_3\,\, e^{-S^{E2{-}D6}}
\end{equation}
After performing the integration over the $\ov \tau$ modes and
$\mu^0$, $\ov \mu^0$ we are left with
\begin{equation}
\int d^4 x d^2\theta  \prod^3_{A=1} \, d^2 \omega d^2\ov \omega\, d
\mu'_A\, d \mu_A \prod^2_{B=1} d  \ov \mu'_B \, d\ov \mu_B \,d \ov
\mu_3 d\mu'_0 d\ov \mu'_0 \,\,e^{-(m_{\ov \mu^A}\, \ov \mu^{A} \,
\mu'_{A} + m_{\mu^B} \,\mu^{B} \, \ov \mu'_{B} )}
\end{equation}
where we omit the term $\vec{D}\cdot \omega \,\vec{\sigma}
\,\bar\omega $ which is irrelevant for the analysis. Next we use the
mass terms for saturating the remaining fermionic charged instanton
modes:
\begin{equation}
\int d^4 x d^2\theta  \prod^3_{A=1} \, d^2 \omega d^2\ov \omega\, d
\ov \mu_3 d\mu'_0 d\ov \mu'_0 \,\,e^{-(m_{\ov \mu_0}\, \ov \mu^0 \,
\mu'_{0} + m_{\mu^0} \,\mu^{0} \, \ov \mu'_{0})}\,\,.
\end{equation}
Note that we cannot saturate the $\mu'_0$ and $\ov \mu'_0$ modes since
we used already $\mu_0$ and $\ov \mu_0$ to saturate the universal $\ov
\tau$ modes. Since there is no other way to soak up these modes the
whole path integral vanishes. Thus we conclude that a generic rigid
$U(1)$ instanton does not contribute to the superpotential, even when
intersects a background D-brane.

\medskip

Let us comment on the saturation of massive modes appearing at
intersections of D-branes and instantons invariant under some
orientifold projection. As mentioned in the introduction, introducing
an orientifold is one of the conventional mechanisms for getting rid
of $\ov\tau$ modes. In this case, due to the absence of the $\ov \tau$
modes we can saturate all fermionic modes via the mass terms apart
from the massless $\mu_3$. As mentioned earlier this mode is the
massless charged mode often called $\lambda$. It can couple to open
string matter fields located between two D6-branes and thus induces
superpotential contributions involving these open string matter
fields.

\medskip

Finally let us briefly compare this setup with the gauge instanton
case discussed in section \ref{sec:gauge-review}, for which a
superpotential contribution can be generated. The crucial difference
between these two configurations is the massiveness of the involved
instanton modes in case of a non-trivial intersection.  We showed that
the naive expectation that the $\mu_A$ and $\ov \mu_A$ pair up to
create a mass term fails, due to the fact that they have different
masses in the regime in which all supersymmetry is broken. In order to
have a proper mass term the presence of additional fermionic modes
$\mu'_A$ and $\ov \mu_A'$ is required, whose presence we confirmed in
section \ref{sec:D-brane-instanton-system} by applying preserved and
broken supersymmetry to the bosonic states $\omega$ and $\ov \omega$
. While the undesired $\ov \tau$ modes can be saturated via the
coupling \eqref{eq:coupling-tau}, in the path integral there is no
coupling to $\mu'$ and $\ov \mu'$ not involving $\mu$ and $\ov \mu$,
respectively. Thus the modes $\mu'$ and $\ov \mu'$ cannot be saturated
and the path integral vanishes.

\section{Some related systems}
\label{sec:dual-pictures}

In the previous discussion we focused on the system of branes
intersecting at angles in type IIA string theory. Using some
well-known duality relations of string theory it is simple to carry
over the lesson from the intersecting case to other corners of the
string theory moduli space. In particular, we will (briefly) discuss
the implications for type IIB string theory. Before going into this,
let us discuss a somewhat subtle point involving the stability of
instantons as we move in moduli space.

\subsection*{Duality and holomorphy of instanton superpotentials}

The issue is the following: generically mirror symmetry will map
smooth geometric compactifications into dual conformal field theories
with no simple geometrical description
\cite{Witten:1993yc,Aspinwall:1993yb,Aspinwall:1993nu}. In order to
obtain a geometric description of the dual we need to move in the dual
K\"ahler moduli space into a region admitting a large volume
description.

One might worry that upon doing this, since the spectrum of BPS
instantons will generically change discontinuously as we move in
moduli space, the results we obtained for intersecting branes are no
longer relevant for the regions of interest. In other words, there
exists the possibility that the non-perturbative superpotential is not
a smooth function in K\"ahler moduli space, and we cannot extrapolate
our vanishing results.

The resolution of this issue comes, as in the vanishing argument given
in the introduction, from holomorphy of the resulting low-energy
superpotential. The non-perturbative superpotential due to instantons
is a holomorphic function in K\"ahler moduli space, and we can
reliably extend the result we obtained in the type IIA description to
the geometric regime of the dual, even if the microscopic description
might change discontinuously. By analyticity this also means that,
since the vanishing of the non-perturbative superpotential holds for
an open subset of the type IIA complex structure moduli space, it must
hold everywhere in the K\"ahler moduli space of the dual.

With these remarks in place, let us proceed to discuss the extension
of our results to type IIB.

\subsection{Branes at singularities in Type IIB}

The first dual picture we want to discuss is that of branes at
singularities in type IIB
\cite{Kachru:1998ys,Klebanov:1998hh,Morrison:1998cs}. For
definiteness, and in order to make contact with some existing
literature, let us take the case of (fractional) D3 branes located at
the $\bC^3/\bZ_2\times \bZ_2$ orbifold. This system was studied using
CFT techniques in \cite{Argurio:2007vqa}. Of particular interest to us
is the discussion in section~3.3 of that paper, where it was argued
that instantons wrapping nodes not wrapped by fractional D3 branes
(the dual of instantons wrapping cycles not homologous to the cycle of
the branes) do not give contributions to the superpotential, agreeing
with our results.

In \cite{Argurio:2007vqa} only the massless modes from the instanton
to the brane were considered, and one might wonder what is the dual of
the tower of massive modes we found before. The answer is that the
spacing of the modes in the tower, which in the intersecting brane
case we took to be small (proportional to the small angles), is at the
quiver point of the order of the string scale. This will become
clearer in next section, when we blow up the singularity. We will
recover a tower of states with arbitrarily small mass separations,
given by the Kaluza-Klein scale of the blown-up exceptional
cycle. This scale is dual to the angle in the type IIA picture, and it
becomes of the string scale at the quiver point.

We can generalize this discussion easily to branes located at
arbitrary toric singularities (abelian orbifolds are particular
examples of toric singularities). The technology to deal with these
cases is well developed by now, in the form of dimer models
\cite{Hanany:2005ve,Franco:2005rj,Hanany:2005ss,Feng:2005gw} (see also
\cite{Kennaway:2007tq,Yamazaki:2008bt} for reviews). Although there
are no available CFT techniques for studying arbitrary toric
singularities, all of them descend from orbifold singularities upon
performing partial resolutions.\footnote{See for example
  \cite{GarciaEtxebarria:2006aq} for efficient techniques to perform
  the resolutions.} These resolutions appear in the theory on the
brane as Higgsings, which cannot generate a non-perturbative potential
for stringy D-brane instantons if there was none in the original
orbifold theory (and Higgsings cannot map gauge instantons into
stringy instantons).

The duality between D3 branes at toric singularities (and the
associated dimer models) and intersecting D6 branes in the mirror of
the toric CY is well understood by now \cite{Feng:2005gw}, and we
refer the curious reader to that paper for details of the duality
map.

\subsection{Magnetized branes in Type IIB/F-theory}

Let us now resolve the singularity by moving away from the quiver
point in K\"ahler moduli space into the large volume regime. This can
be done most systematically for singularities which can be resolved by
blowing up a four cycle; one important example of this class are the
complex cones over the del Pezzo surfaces. For instance, blowing up
the $\cN=1$ $\bC^3/\bZ_3$ orbifold gives the complex cone over $dP_0$.

When this is the case, we have a rather powerful large volume
description of the physics in terms of exceptional collections
\cite{Cachazo:2001sg,Wijnholt:2002qz,Aspinwall:2004vm,Herzog:2005sy,Bergman:2005kv}
living on the blown up four-cycle. Physically we can interpret these
exceptional collections as a nice basis of fractional D3 branes, which
at large volume are interpreted as D7 (or $\overline{D7}$) branes with
fluxes on their worldvolume. We can then map the fractional branes
discussed in the previous section to elements in the exceptional
collection by using the $\Psi$ map proposed in
\cite{Hanany:2006nm}.\footnote{For completeness, let us mention that
  the full physical interpretation of the large volume objects one
  obtains from the fractional branes is slightly more complicated:
  they are objects of the derived category of coherent sheaves
  $D^\flat(X)$, where $X$ is the Calabi-Yau cone, see for example
  \cite{Aspinwall:2004jr} for a detailed discussion.}

Using these ideas we can follow the duality between the intersecting
branes and the mirror large volume branes in detail. The Kaluza-Klein
tower of states (going from the instanton to the brane) in the large
volume picture is dual to the tower of states spaced by the brane
intersection angle analyzed in section~\ref{sec:cft}, and the whole
discussion done there applies with little change to the system at
hand.

Thus, this Type IIB mirror description of the system studied in
section \ref{sec:cft} does not seem to give a non-perturbative
superpotential; it would be interesting to address these aspects
within a fully F-theoretical context.

\section{Non-commutative instantons}
\label{sec:noncommutative}

In the introduction we discussed an argument based on the
considerations of
\cite{GarciaEtxebarria:2007zv,GarciaEtxebarria:2008pi} why the system
with instantons and branes at angles should not contribute to the
superpotential. The key observation was that the instanton can
misalign from the brane without recombining, so it cannot contribute
to the superpotential when misaligned, and by holomorphy of the
superpotential on the closed string moduli that create the
misalignment, it should not contribute to the superpotential when
aligned either.

We wish to emphasize in this section the importance of the condition
that the instanton should not be able to recombine with the brane. As
verified explicitly in section~\ref{sec:cft}, the spectrum of bosonic
strings going from the instanton to the brane at an angle in the
internal space has positive nonvanishing mass in the misaligned
regime, so no recombination is possible. In this section we would like
to discuss the non-commutative instanton in string theory as a simple
example in which the instanton {\em does} dissolve into the brane when
misaligned, due to the appearance of tachyons.

\medskip

Let us consider a single D$(p+3)$ brane extended in the four Minkowski
directions and wrapping a supersymmetric $p$-cycle $\Sigma_p$ in the
internal space, such that at low energies the theory on the
worldvolume of the brane reduces to $\cN=1$ SYM with gauge group
$U(1)$. We are interested in non-perturbative corrections to the
superpotential coming from euclidean D$(p-1)$ branes wrapping
$\Sigma_p$. As discussed in
\cite{Aganagic:2007py,GarciaEtxebarria:2007zv,Petersson:2007sc,GarciaEtxebarria:2008iw},
even though in $U(1)$ gauge theories there are no instanton effects,
in string theory the D-brane instanton wrapping the same node as the
brane does generate a low energy superpotential:
\begin{equation}
  W_{eff} = e^{-S_{inst}}= e^{-\frac{1}{g_s}V\{\Sigma_p\}}
\end{equation}
with $V\{\Sigma_p\}$ the complexified volume of $\Sigma_p$.  It is
possible to misalign the brane and the instanton by switching on a
field strength on the Minkowski part of the D$(p+3)$ brane, or
equivalently by switching on a $B$-field.

There is a well known dual description of such a system:
non-commutative gauge theory \cite{Seiberg:1999vs}. We will see here
that the result one obtains from non-commutative gauge theory nicely
matches what one requires from holomorphy, and in particular a tachyon
appears when the instanton becomes misaligned with the D-brane.

\medskip

For completeness, we include here a telegraphic review of the main
points we will need. We refer the reader to
\cite{Nekrasov:2000ih,Harvey:2001yn} for some nice set of lectures on
non-commutative gauge theories, and to \cite{Seiberg:1999vs} for the
original embedding of non-commutative field theories in string theory
that we will be using.\footnote{There were interesting previous
  embeddings of non-commutative gauge theory in string theory, see for
  example
  \cite{Connes:1997cr,Berkooz:1998st,Nekrasov:1998ss}.}

In a non-commutative gauge theory the space-time coordinates satisfy
the algebra:
\begin{equation}
  [x_\mu,x_\nu] = i \theta_{\mu \nu}
\end{equation}
with $\theta_{\mu \nu}$ some antisymmetric tensor. It will prove
convenient to split $\theta$ into its self-dual and anti-self-dual
parts:
\begin{equation}
  \theta_{\mu \nu} = \xi_{+}^i \eta_{\mu\nu}^{i} + \xi_{-}^i \bar\eta_{\mu\nu}^i
\end{equation}
with $\eta$ and $\bar\eta$ the self-dual and anti-self-dual 't Hooft
symbols respectively (see appendix~A of \cite{Dorey:2002ik} for
explicit expressions). The index $i$ runs from 1 to 3.

The way we engineer such a deformation of the gauge theory in our
stringy setup is simple: we just need to include a $B$-field in the
four Minkowski directions \cite{Nekrasov:1998ss,Seiberg:1999vs}. The
non-commutativity parameter will then be given by $\theta=B^{-1}$.

Let us discuss the supersymmetries preserved by such a background. We
pick the convention that the D$(p-1)$ instanton preserves the same
supersymmetry as a self-dual field strength on the D$(p+3)$ brane. By
the same token, if we want to misalign the instanton with respect to
the brane, we should switch on the anti-self-dual part of the field
strength. In terms of $\theta$, we are thus interested in switching on
$\xi_{-}$.

It is a well known fact that switching on $\xi_{+}$ does not modify
the moduli space of instantons for a $U(1)$ theory (see
\cite{Seiberg:1999vs}, for example). In particular, it does not remove
the small instanton singularity corresponding to the stringy D$(p-1)$
instanton. In contrast, switching on the anti-self-dual term $\xi_{-}$
has a drastic effect on the moduli space: it removes the small
instanton singularity.

Let us briefly recall how this modification of the moduli space comes
about. The introduction of $\theta$ modifies the ADHM construction of
instantons by the introduction of a Fayet-Iliopoulos term in the
bosonic ADHM constraints:
\begin{equation}
  \label{eq:nc-ADHM}
  \ov\omega \sigma^i \omega = \xi_{-}^i
\end{equation}
Since the radius $\rho$ of the instanton is given by $\rho=\ov\omega
\omega$, this implies that the instanton can no longer have zero size,
and it necessarily dissolves into the brane, as advertised.

It is not hard to see the required tachyon explicitly, either. Let us
write
\begin{equation}
  \omega = \begin{pmatrix}\phi_1 \\
    \phi_2
    \end{pmatrix}
\end{equation}
and choose $\xi_{-}^i = (0,0,\xi)$. From (\ref{eq:nc-ADHM}), we have:
\begin{equation}
  |\phi_1|^2 - |\phi_2|^2 = \xi
\end{equation}
Equivalently, recalling that (\ref{eq:nc-ADHM}) is actually a D-term
in the instanton action, we have:
\begin{equation}
  S_{inst} = \ldots + \left(  |\phi_1|^2 - |\phi_2|^2 - \xi \right)^2
\end{equation}
Note how for any nonzero $\xi$ one of $\phi_1,\phi_2$ gets a negative
mass squared, becoming the required tachyon. The condensation of the
tachyon restores supersymmetry, and allows the (diluted) instanton to
contribute.

\section{Conclusions}

In this paper we have discussed in some detail the system of branes
and instantons intersecting at angles, with the conclusion that they
cannot contribute to the superpotential. We gave both macroscopic
arguments based on holomorphy (in the introduction), and microscopic
arguments using a local CFT analysis (section~\ref{sec:cft}). In
section~\ref{sec:dual-pictures} we discussed some important
implications of our result for the dual type IIB picture, and in
section~\ref{sec:noncommutative} we briefly discussed how the presence
of tachyons allows for non-perturbative contributions even for
instantons that can be misaligned, which we illustrated in the
interesting case of non-commutative instantons.

\medskip

We believe our result to be robust, and widely applicable for any
system where the instanton has a supersymmetry phase independent of
the brane (following the arguments of
\cite{GarciaEtxebarria:2008pi}). We would like to emphasize at this
point that the key issue is really the existence or not of tachyons at
every point of moduli space where the instanton gets misaligned with
respect to the background branes. In general this is a condition that
can be easily checked. Modes with positive mass have no effect on the
saturation of $\ov\tau$, as we showed in detail in
section~\ref{sec:cft}. Microscopically we could also see this from the
requirement of supersymmetry of the instanton effective action, which
disallowed massive modes from saturating massless modes.

\medskip

Possibly the most conspicuous omission in the analysis above is that
of the massive modes of string scale. That is, modes with masses of
the form:
\begin{equation}
  m = c + \sum_{i=1}^3 a_i \theta_i
\end{equation}
with $c\neq 0$. If such an effect were present, it would be in fact
the leading contribution to the otherwise vanishing couplings in the
superpotential. We do not expect such modes to alter the conclusion
for the following reasons.  As mentioned in the introduction, a
similar structure of higher massive modes is present in the gauge
instanton case, where it is known that they do not change the
vanishing results one obtains by looking at the massless sector
only. Another argument follows from the world-sheet charge
conservation considerations in section~\ref{sec:cft}: in order to
construct a mass term we need two fermionic modes, but by world-sheet
charge conservation arguments only one of them can couple to
$\ov\tau$.

\medskip

Furthermore, one might also entertain the idea that $g_s$ effects
might play a role in the lifting of $\ov\tau$ zero modes in the
intersecting instanton case. One strong argument against this
possibility is the holomorphy argument which still holds in the low
energy supergravity: if one can misalign the instanton with the brane
no superpotential should be generated.

\medskip

Another interesting avenue of research concerns higher F-terms
\cite{Beasley:2004ys,Beasley:2005iu}. In general, when $\ov\tau$ modes
are not saturated BPS instantons contribute to higher fermionic
F-terms \cite{Blumenhagen:2007bn}, and here intersecting brane
instantons can have an important effect. In particular, they will give
interesting deformations of the moduli space of the string
compactification.

\medskip

We hope that this work helps to clarify part of the global structure
of non-perturbative effects in string theory compactifications, a
fascinating topic.

\acknowledgments

We would like to acknowledge interesting discussions with Ralph
Blumenhagen, Jim Halverson, Liam McAllister, Mike Schulz and Angel
Uranga. We are especially grateful to Timo Weigand for discussions and
comments on the manuscript. We would also like to thank Jonathan
Heckman, Joseph Marsano, Natalia Saulina, Sakura Sch{\"a}fer-Nameki
and Cumrun Vafa for extensive discussions after the first version of
this paper appeared. I.G.E. thanks Nao Hasegawa for kind support and
constant encouragement. This work is supported by DOE grant
DE-FG05-95ER40893-A020, NSF RTG grant DMS-0636606 and Fay R. and
Eugene L. Langberg Chair.

\bibliographystyle{JHEP}
\bibliography{refs}

\end{document}